\documentclass[12pt]{article}
\pdfoutput=1
\usepackage{amssymb,amsmath}
\usepackage{epsfig}
\usepackage{epstopdf}
\usepackage{latexsym}
\usepackage{graphicx}
\usepackage{subfigure}
\usepackage{booktabs}
\usepackage{bbm}
\usepackage[mathscr]{euscript}
\usepackage[margin=20pt,small]{caption}
\let\originaleqref\eqref
\renewcommand{\eqref}{\originaleqref}
\usepackage{bigints}
\usepackage{color}
\usepackage{ifthen}
\usepackage{hyperref}
\newboolean{showToPublic}
\setboolean{showToPublic}{true}
\newcommand{\todocmd}[2]{{{#1}}{{#2}}}
\newcommand{\todo}[2]{\ifthenelse {\boolean{showToPublic}} {\todocmd{#1}}{\todocmd{#2}}}

\def\f{\phi_1}
\def\fp{\phi_1'}
\def\fpp{\phi_1''}

\def\cN{{\cal N}}
\def\cO{{\cal O}}

\def\cL{{\cal L}}

\def\cL{{\cal L}}
\def\cA{{\cal A}}

\def\cU{{\cal U}}

\def\wt{\widetilde}
\def\mk{\mathfrak}

\definecolor{cardinal}{rgb}{0.6,0,0}
\definecolor{darkgreen}{rgb}{0,0.5,0}
\definecolor{golden}{rgb}{0.92, 0.7, 0}
\definecolor{midnight}{rgb}{0, 0, 0.5}
\definecolor{darkblue}{rgb}{0.2, 0, 0.8}

\let\k=\kappa

\newcommand{\be}{\begin{equation}}
\newcommand{\ee}{\end{equation}}
\newcommand{\bea}{\begin{eqnarray}}
\newcommand{\eea}{\end{eqnarray}}

\begin{document}

\begin{titlepage}

\begin{flushright}
UTTG-22-14, TCC-024-14\\
CCTP-2014-24, CCQCN-2014-49
\end{flushright}

\bigskip
\centerline{\Large \bf Weak Field Collapse in AdS: Introducing a Charge Density }
\bigskip
\bigskip
\centerline{{\bf Elena Caceres$^{1,2}$, Arnab Kundu$^{3,4}$, Juan F.~Pedraza$^{2,5}$ and Di-Lun Yang$^{6,7}$}}
\bigskip

\centerline{$^1$ Facultad de Ciencias, Universidad de Colima,}
\centerline{Bernal Diaz del Castillo 340, Colima, Mexico.}
\vspace{2mm}
\centerline{$^2$ Theory Group, Department of Physics \&  Texas Cosmology Center }
\centerline{The University of Texas at Austin, Austin, TX 78712, USA.}
\vspace{2mm}
\centerline{$^{3}$ Department de F\'{i}sica Fondamental \& Institut de Ci\`{e}ncies del Cosmos,}
\centerline{Universitat de Barcelona, Mart\'{i} i Franqu\`{e}s 1, E-08028 Barcelona, Spain.}
\vspace{2mm}
\centerline{$^{4}$ Saha Institute of Nuclear Physics}
\centerline{1/AF Bidhannagar, Kolkata 700064, India.}
\vspace{2mm}
\centerline{$^{5}$ Perimeter Institute for Theoretical Physics}
\centerline{Waterloo, Ontario N2L 2Y5, Canada.}
\vspace{2mm}
\centerline{$^6$ Crete Center for Theoretical Physics, Department of Physics}
\centerline{University of Crete, 71003 Heraklion, Greece.}
\vspace{2mm}
\centerline{$^7$ Department of Physics, Chung-Yuan Christian University}
\centerline{Chung-Li 32023, Taiwan.}
\bigskip

\begin{abstract}

\noindent We study the effect of a non-vanishing chemical potential on the thermalization time of a strongly coupled large $N_c$ gauge theory in $(2+1)$-dimensions, using a specific bottom-up gravity model in asymptotically AdS space. We first construct a perturbative solution to the gravity-equations, which dynamically interpolates between two AdS black hole backgrounds with different temperatures and chemical potentials, in a perturbative expansion of a bulk neutral scalar field. In the dual field theory, this corresponds to a quench dynamics by a marginal operator, where the corresponding coupling serves as the small parameter in which the perturbation is carried out. The evolution of non-local observables, such as the entanglement entropy, suggests that thermalization time decreases with increasing chemical potential. We also comment on the validity of our perturbative analysis.

\end{abstract}
\bigskip

\end{titlepage}

\newpage

\tableofcontents

\setcounter{equation}{0}
\section{Introduction}

The AdS/CFT correspondence describes an equivalence between a classical gravitational dynamics and a large $N_c$ gauge theory. This remarkable correspondence has proved to be very useful in addressing aspects of strongly coupled dynamics in various models, ranging from understanding aspects of strongly coupled Quantum chromodynamics (QCD) to condensed matter-inspired systems. See {\it e.g.}~\cite{CasalderreySolana:2011us, Hartnoll:2009sz} for recent reviews on some of these attempts.

Most of such endeavours usually discuss equilibrium properties of a class of strongly coupled SU$(N_c)$ gauge theories at large $N_c$. However, since this is a correspondence between the path integrals of the corresponding Quantum Field Theory (QFT) and the classical Gravity description, it is natural to assume that it extends to time-dependent dynamical situations as well. In fact, dynamical processes in a prototypical field theory model are extremely interesting to explore and learn about, since we do not have a good understanding of the governing rules and laws for systems completely out-of-equilibrium, specially at strong coupling. In this regime, AdS/CFT correspondence is potentially a very useful tool.

Such time-dependent issues fall under two broad categories: the study of quantum quenches, where a system is prepared in an energy eigenstate of a given Hamiltonian. The Hamiltonian is then perturbed by a time-dependent external parameter. Recent developments in cold atoms experiments have initiated a very active research where this perturbation occurs abruptly, see {\it e.g.}~\cite{quench_bib} for a condensed-mater approach and \cite{Das:2011nk, Basu:2011ft, Das:2010yw} for a review of the holographic attempts. The other broad issue is to understand the physics of thermalization for strongly coupled system. See {\it e.g.}~\cite{Danielsson:1999zt, Danielsson:1999fa, Giddings:2001ii} for earlier works on this. More recently, there has been a renewed interest to understand the physics of thermalization at strong coupling to shed light on the physics of the quark-gluon plasma (QGP) at the Relativistic Heavy Ion Collider (RHIC) and the Large Hadron Collider (LHC). Most of these works rely heavily on numerical efforts that study a black hole formation process in an asymptotically anti de-Sitter (AdS) space, see {\it e.g.}~\cite{Chesler:2008hg, Chesler:2009cy, Heller:2011ju, Heller:2012je, Heller:2012km, Casalderrey-Solana:2013aba, vanderSchee:2013pia,Casalderrey-Solana:2013sxa,  Balasubramanian:2013rva, Balasubramanian:2013oga, Craps:2013iaa, Cardoso:2013vpa, Fernandez:2014fua}.

A generic thermalization process typically describes the dynamical evolution from a ``low temperature" phase to a ``high temperature" thermal state, where the evolution is highly non-trivial and, in the case of a black hole formation process, highly non-linear as well. Holographically, such a process in asymptotically AdS space can be set up by turning  on a non-normalizable (or a normalizable) mode of a bulk scalar field; as  a  result a shell of the corresponding field collapses in AdS space. It was shown in \cite{Bhattacharyya:2009uu}, using a weak-field perturbation method, that if the boundary non-normalizable mode is chosen to be coordinate independent and only have support over a brief time interval, the collapse of the corresponding homogeneous wave will always lead to black hole formation.\footnote{If the non-normalizable mode is not turned on, there is a large class of homogeneous, spherically symmetric, initial conditions whose temporal evolution does not lead to black hole formation \cite{Bizon:2011gg,Balasubramanian:2014cja,Buchel:2014dba}. These bulk solutions correspond to states of the boundary conformal field theory (CFT) that fail to thermalize at late times. These solutions have recently been proposed as the analogue-dual of the ``quantum revivals'' observed in \emph{finite} size isolated quantum systems and widely studied in the condensed matter literature \cite{Abajo-Arrastia:2014fma}.}

On the other hand, an alternate approach is to {\it phenomenologically model} the black hole formation process with as much simplicity as possible, such that the corresponding geometry can be probed to learn further physics. The hope is to learn at least qualitatively useful lessons which are presumably not heavily dependent on the details of the model. In the present context this can be achieved by exploring the AdS-Vaidya background, which describes a smooth evolving geometry from an empty AdS to an AdS-Schwarzschild background. Gravitationally, this geometry describes the collapse of a null dust in an asymptotically AdS-space. Probing this geometry has already led to numerous interesting results, see {\it e.g.}~\cite{AbajoArrastia:2010yt, Balasubramanian:2010ce, Balasubramanian:2011ur, Albash:2010mv, Galante:2012pv, Caceres:2012em, Caceres:2012px}, where the behavior of various non-local observables in such a dynamical geometry has been explored.

In \cite{Bhattacharyya:2009uu}, an interesting bridge between these two approaches has been established. The authors studied a collapse process for a massless scalar field in a so called ``weak field approximation" limit, where the amplitude of the perturbation was chosen small and a perturbative solution of the Einstein's equations was obtained. At the leading order, this solution takes the form of an AdS-Vaidya background which is characterized by one mass function that interpolates between an AdS vacuum to an AdS-Schwarzschild geometry. In the dual field theory side, this corresponds to a dynamical evolution from zero temperature ground state to a thermal state of a certain large $N$ gauge theory (such as the ${\cal N}=4$ super Yang-Mills or the ABJM theory). Thus, at least at the leading order, the analysis of \cite{Bhattacharyya:2009uu} validates the AdS-Vaidya-based phenomenology from a first principle gravity computation, see \cite{Garfinkle:2011tc} for more details.

Motivated by this observation, we will explore the possibility of introducing a conserved charge in the boundary theory --- the simplest case of an U$(1)$-charge in this article --- in an analogue of the ``weak field approximation" limit starting from an effective gravity action. Our motivation is to address how thermalization time is affected in the presence of global conserved charges in a strongly coupled system, with as much analytical control as possible. This may be relevant for understanding the effect of a non-vanishing chemical potential (or a finite-density) on the strong coupling dynamics of out-of-equilibrium QCD at RHIC or at LHC.\footnote{We should note here that QGP physics might not be the best example here since both at the RHIC\cite{Andronic:2005yp} and at the LHC\cite{Stachel:2013zma}, the baryon chemical potential is sufficiently low.} We will thus generalize the construction of \cite{Bhattacharyya:2009uu} introducing a chemical potential, which --- albeit in a suitable approximation --- will provide an analytical control on the background. This also provides us with a model in which non-local observables can subsequently be studied to explore the behaviour of thermalization time, in the spirit of \cite{Galante:2012pv, Caceres:2012em}.

Our results may be of general interest, beyond the QGP physics. For example, a qualitative behaviour, if it is universal, may shed light in condensed matter systems which are typically accompanied by a non-vanishing chemical potential. Moreover, to the best of our knowledge, there is no known field theoretic result about how thermalization time scales in a strongly coupled finite density system. Thus it is useful to explore models where this possibility can be realized.

This article is divided in the following parts: We begin with a summary of our results and a brief discussion on those in the next section. In section \ref{sec:level3} we present the effective gravity model and provide the details of the perturbative solution. We then discuss the initial and the final states in details in section \ref{sec:level4}, and subsequently discuss the behaviour of the thermalization time in section \ref{sec:level6}. Finally we conclude in section \ref{sec:level7}.

\setcounter{equation}{0}
\section{Summary and Discussion\label{sec:level2}}

Let us begin with a more specific description. We will discuss a thermal quench, and for simplicity we will restrict ourselves to a $(2+1)$-dimensional large $N_c$ gauge theory in the strong coupling regime. Now, consider the Hamiltonian (or the Lagrangian) of the system, denoted by $H_0$ (or $\cL_0$), which is perturbed by a time-dependent perturbation of the form
\begin{eqnarray}
H _{\lambda} = H_0 + \lambda(t) \delta H_\Delta  \quad \implies \quad \cL_\Delta = \cL_0 + \lambda(t) \cO_\Delta \ ,
\end{eqnarray}
where $H_0$ (or $\cL_0$) describes the Hamiltonian (or the Lagrangian) of the original quantum field theory, $\lambda(t)$ is an external parameter and $\delta H_\Delta$ (or $\cO_\Delta$) corresponds to the deformation of the QFT by an operator of dimension $\Delta$.

Here we will restrict ourselves completely on asymptotically locally AdS-spaces, which implies that there is an underlying conformal field theory (CFT) that governs the physics. As a first attempt, we restrict ourselves to the case of $\Delta = d$, which here becomes $\Delta = 3$, {\it i.e.}~an exactly marginal deformation. In principle, it is possible to study the quench dynamics by a relevant operator as well. However, we know that relevant operators can trigger an RG-flow and there may be a new CFT --- or perhaps a non-relativistic cousin of it --- that emerges in the infrared. In such a situation, unless we consider a temperature scale much larger than the scale set by the relevant operator, the black hole formation process may be governed by this infrared geometry instead.\footnote{Note that, quench dynamics of relevant operators have been considered in details in {\it e.g.}~\cite{Buchel:2012gw,Buchel:2013lla,Buchel:2014gta}.}

To avoid this possible subtlety for relevant operators, we will consider an exactly marginal operator, which does not require a hierarchy between the RG-scale and the temperature-scale. Thus the underlying CFT will remain the same and in the gravitational dual it will suffice for us to specify the asymptotically locally AdS condition with a given radius of curvature as the boundary condition.\todo{}{\color{red}[AK: mention that this cannot be embedded in ABJM, because all scalars have a non-zero mass there.]}

Now we need to specify the initial conditions. Typically this is specified at $t \to -\infty$ (in which limit $\lambda \to 0$) as a particular energy eigenstate of the QFT Hamiltonian $H_0$. As the new coupling is turned on, depending on whether the process is adiabatic or abrupt, the system is expected to evolve differently with the Hamiltonian $H_\lambda$. In a quantum mechanical system, under an adiabatic process the system remains in an energy eigenstate whose energy evolves with time following the response of the time-evolution of $\lambda(t)$. On the other hand, for an abrupt quench, the system evolves in a linear superposition of eigenstates of the new Hamiltonian $H_\lambda$. \todo{}{\color{red}[AK: Can we make this more precise in our holographic results?]}  Here we will focus on the fast quench only\footnote{We will momentarily explain what we mean by a ``fast quench".}, in which the initial state is macroscopically characterized by $\left\{E, \langle \cO_\phi \rangle , \mu, T \right \}_{\rm initial}$ --- with $E$, $\langle \cO_\phi \rangle$, $\mu$ and $T$ respectively representing the energy, VEV for the marginal operator, chemical potential and the temperature of the state in consideration --- and the final state is macroscopically characterized by $\left \{E, \langle \cO_\phi \rangle , \mu, T \right \}_{\rm final}$.

To properly account for the scale of measuring dimensionful quantities, we define the following dimensionless parameters:
\begin{eqnarray}
\kappa_E = \frac{E}{T^3} \ , \quad \kappa_{\langle \cO_\phi \rangle} = \frac{\langle \cO_\phi \rangle}{T^3} \ ,  \quad \kappa_\mu = \frac{\mu}{T} \ ,  \label{dimlesspara1}
\end{eqnarray}
which means that we measure all dimensionful quantities in units of temperature of the corresponding state. Evidently, there can be several hierarchy of scales depending on how $\kappa_E$, $\kappa_{\langle \cO_\phi \rangle}$ and $\kappa_{\mu}$ are parametrically separated. Furthermore, there are a couple of dimensionful parameters associated with the quench process itself: the energy injected (denoted by $\Delta E$), and the duration (denoted by $\delta t$) or the rapidity of the quench. Thus we can further define
\begin{eqnarray}
\kappa_{\rm quench} = \left(\Delta E \right) \left (\delta t\right)^3 \quad {\rm with} \quad \Delta E = E_{\rm final} - E_{\rm initial} \ .  \label{dimlesspara2}
\end{eqnarray}

Now, our initial state is characterized by the following parametric regime
\begin{eqnarray}
\kappa_E^{\rm initial} \sim \cO(1) \ , \quad  \kappa_{\langle \cO_\phi \rangle}^{\rm initial} = 0 \ , \quad  \kappa_\mu^{\rm initial} \ll 1 \ .  \label{condini}
\end{eqnarray}
The quench process, to be amenable to a perturbative analysis, is characterized by
\begin{eqnarray}
\kappa_{\rm quench} \ll 1 \quad {\rm with} \quad  \cO\left(\kappa_{\rm quench} \right) = \cO\left(\kappa_{\mu}^{\rm initial} \right) \ .  \label{conddyn}
\end{eqnarray}
Finally, the final state is characterized by
\begin{eqnarray}
\kappa_E^{\rm final} \sim \cO(1) \ , \quad  \kappa_{\langle \cO_\phi \rangle}^{\rm final} \sim \kappa_{\mu}^{\rm initial}  \ll 1  \ , \quad  \kappa_\mu^{\rm final} \sim \kappa_{\mu}^{\rm initial} \ll 1 \ .  \label{condfin}
\end{eqnarray}
We remind the reader that the conditions in (\ref{condini})-(\ref{condfin}) are specific to our perturbative analysis. We further note that, the perturbative solution that we construct is not similar to the AdS-RN-Vaidya geometry described and analyzed in \cite{Caceres:2012em} and thus provides a more generic case-study. It is amusing to further note that the regime of parameters outlined in (\ref{condfin}) physically implies that we are considering a small chemical potential limit, which is qualitatively similar to the QGP-phase at the LHC.\todo{}{\color{red}[AK: confirm this, and make it more quantitative.]}

The geometric data describing the corresponding evolution is given by a metric, a gauge field, and a scalar field: $\{G, A, \phi \}$ of the following form
\begin{eqnarray}
ds^2 & = & G_{\mu\nu} \left(\cU_{\rm EF} \right) dX^\mu dX^\nu \ , \nonumber\\
         & = & -\frac{2}{z^2} dv dz - g(z,v) dv^2 + f(z,v)^2 \sum_{i=1}^2 dx_i^2 \ , \label{metricsum} \\
      A & = & A_v (z, v) dv \ , \quad \phi = \phi(z, v) \ ,
\end{eqnarray}
where $\cU_{\rm EF}$ denotes the ingoing Eddington-Finkelstein (EF) patch, $g(z, v)$ and $f(z,v)$ are two functions that are determined by solving the equations, $z$ is the radial coordinate (in which the boundary is located at $z \to 0$) and $v$ is the EF-ingoing null direction. \todo{}{\color{red} [AK: present a pictorial form of the horizon formation, comprising of the following things: the old horizon, the apparent horizon and the new horizon. etc.]} We obtain a dynamical evolution that can be summarized as:
\begin{eqnarray}
\left. \left \{G, A, \phi  \right \} \right |_{v \to - \infty} & = & {\rm AdS-RN}_{\rm initial} \left(\mu_i , T_i,  \phi =0 \right) \ , \\
\left. \left \{G, A, \phi  \right \} \right |_{v \gg \delta t} & = &  {\rm AdS-BH}_{\rm final} \left( \mu_f, T_f, \phi \not = 0 \right) \ .
\end{eqnarray}
Let us now comment on the evolution of the event-horizon and the apparent horizon. We begin with the notion of the apparent horizon. Following \cite{Caceres:2012em}, let us define the tangent vectors to the ingoing and outgoing null geodesics
\begin{eqnarray}
\ell_{-}  = - \partial _z \ , \quad \ell_{+} = - z^2 \partial_v + \frac{g(z,v)}{2} z^4 \partial_z \ ,
\end{eqnarray}
which satisfy
\begin{eqnarray}
\ell_{-} \cdot \ell_{-} = 0 \ , \quad \ell_{+} \cdot \ell_{+} = 0 \ , \quad \ell_{-} \cdot \ell_{+} = - 1 \ .
\end{eqnarray}
The co-dimension two spacelike surface, which is orthogonal to the tangent vectors above, has the following volume element
\begin{eqnarray}
\Sigma = f\left(z, v\right)^2 \ .
\end{eqnarray}
The expansion of this volume element along the ingoing and outgoing null directions are given by
\begin{eqnarray}
\theta_{\pm} = \cL_{\pm} \log \Sigma = \ell_{\pm}^\mu \partial_\mu \left( \log\Sigma\right) \ .
\end{eqnarray}
Here $\cL_{\pm}$ denotes the Lie derivatives along the null direction corresponding to $\ell_{\pm}^\mu$. Now, we can define the invariant expansion by $\Theta = \theta_{+} \theta_{-}$ which is given by
\begin{eqnarray}
&& \Theta  =  \frac{\left( \partial_z f(z,v)\right)}{f(z,v)} \left[ 2 z^2 \frac{\left( \partial_v f(z,v)\right)}{f(z,v)} - z^4 \frac{\left( \partial_z f(z,v)\right)}{f(z,v)} g(z, v) \right] \ , \quad {\rm with} \nonumber\\
&& \Theta \left( z = z_{\rm aH}\right)  =  0 \ .\label{eq:ahdef}
\end{eqnarray}
Here $z_{\rm aH}$ denotes the location of the apparent horizon.

On the other hand, the event horizon is a null surface in the background (\ref{metricsum}):
\begin{eqnarray}
\cN = z - z_{\rm H}(v) \quad {\rm obeying} \quad G^{\mu\nu} \left( \partial_\mu \cN \right) \left( \partial_\nu \cN \right) = 0 \ ,
\end{eqnarray}
which gives
\begin{eqnarray}\label{eq:ehdef}
\partial_v z_{\rm H}(v) = - \frac{1}{2} z_{\rm H}^2 g \left(z_{\rm H}, v \right) \ .  \label{eq:ehdef}
\end{eqnarray}
Thus, solving (\ref{eq:ahdef}) and (\ref{eq:ehdef}) gives the time-evolution of the apparent and the event horizon, respectively. Evidently, at the initial and the final states they coincide: $z_{\rm H} = z_{\rm aH}$. During the evolution, since the collapse is sourced by a physically reasonable matter field, the apparent horizon lies behind the event horizon, {\it i.e.}~$z_{\rm aH} > z_{\rm H}$ in our choice of the radial coordinate. One way to summarize our perturbative solution is to state that the evolution of {\it e.g.}~the event-horizon can be obtained in a series expansion as follows:
\begin{eqnarray}
{\rm Given \, \, that} \quad g(z, v) & = & \sum_{n=0} \varepsilon^{2n} g_{(2n)} \left(z_{\rm H}, \frac{v}{\delta t} \right)\ , \nonumber \\
{\rm construct} \quad z_{\rm H} & = & z_{\rm H}^{(0)} \left[ 1 + \sum_{n=1} \varepsilon^{2n} \Upsilon_{(2n)} \left(\frac{v}{\delta t}\right) \right] \ ,  \label{pertzH}
\end{eqnarray}
where $\Upsilon_{{2n}}$ can be determined from a first order differential equation of the form
\begin{eqnarray}
\partial_v \Upsilon_{(2n)} = \Xi \left[ \left\{\Upsilon_{(2n)} \right\}, \left\{ \partial_{z_{{\rm H}}}^n g_{(2m)} \right\} \right] \ ,
\end{eqnarray}
where $\Xi$ is a functional of
\begin{eqnarray}
\left\{\Upsilon_{(2n)} \right\} = \left\{\Upsilon_{(2p)} \,  | \, 1 \le p \le n \right\} \ , \quad \left\{ \partial_{z_{{\rm H}}}^n g_{(2m)} \right\} = \left\{ \partial_{z_{{\rm H}}}^p g_{(2q)} \, | \, 0 \le (p,q) \le n \right\} \ ,
\end{eqnarray}
and, finally, $\varepsilon \sim \kappa_{\rm quench}^{1/2} \ll 1$. In section \ref{sec:level6}, we will present a pictorial representation of how the apparent and the event-horizons evole.

It is clear from (\ref{pertzH}) that the asymptotic series captures the physics as long as $|| \varepsilon^{2n} \Upsilon_{(2n)} ||~\ll~1$. It will be shown In section \ref{sec:level5}, that this imposes a constraint and our perturbative analysis is valid up to a time-scale $ t_{\rm pert} \sim 1 / \left( \Delta E \right)^{1/3}$. For $t \gg t_{\rm pert}$, we will need to solve the system of equations numerically, which we will not pursue here. Written in terms of the duration of the pulse, the perturbative treatment is valid up to a time-scale $t_{\rm pert}$ which is given by
\begin{eqnarray}
t_{\rm pert} = \cO \left( \frac{\delta t}{\varepsilon^{2/3}} \right) \ .
\end{eqnarray}
Thus, by tuning $\varepsilon \ll 1$, we parametrically separate $t_{\rm pert}$ and $\delta t$ at will, and in this sense our approach is equivalent to considering an ``fast quench". In the strict fast quench limit, {\it i.e.}~$t_{\rm pert}/ (\delta t) \to \infty$, the $t > t_{\rm pert}$ dynamics is completely frozen and we are left with the perturbative solution for all times.

To measure thermalization time, we need to identify suitable observables, which primarily fall in two classes: local and non-local. Note that, unlike the Vaidya-construction, for gauge-invariant local observables thermalization is not instantaneous in this case.\footnote{We thank Alex Buchel for pointing this to us.} This is buried in the details of the solution in (\ref{metricsum}), since as time varies, the scalar field dynamically acquires a non-zero expectation value. However, such local operators thermalize over a time-scale of $\cO (\delta t )$, and does not contain the information about long-range correlations. On the other hand, non-local operators provide a more global information and we will explore the evolution of entanglement entropy in this article.

Now, let us comment on a na\`{i}ve expectation about the scaling of the thermalization time. It is known that for integrable systems, which contains an infinite number of conserved charges, thermalization does not happen. \todo{}{\color{red}[AK: reference?]} This is intuitively clear, since it becomes unlikely to populate the entire phase space. Furthermore, for a single U$(1)$-charge, if we consider a bosonic system, increasing the chemical potential will enhance Bose-Einstein condensation and thus will inhibit thermalization. For fermionic degrees of freedom, a higher value of chemical potential is associated with a higher Fermi surface which will subsequently need to be populated to achieve a thermal state. Thus, introducing a conserved charge is likely to have a slowing down of the thermalization time.

However, contrary to the expectation outlined in the above paragraph, we gather evidence that thermalization speeds up for increasing chemical potential in the regime $\mu / T \ll 1$. Thus, modulo the caveats of an effective gravity description and the approximate measures of the thermalization time, we are lead to think that the strong coupling dynamics perhaps gives rise to qualitatively new physics. Furthermore, since the thermalization time is unlikely to vanish for arbitrarily high chemical potential, we expect it to either saturate or turn back. In both cases, the scaling of the thermalization time seems to group the dynamics in two qualitatively different regimes: $\mu/ T \ll 1$ and $\mu / T \gg 1$, much like what was observed in \cite{Caceres:2012em}. Now we will turn to discussing the details of our model.

 \setcounter{equation}{0}
\section{Einstein-Maxwell-Dilaton system\label{sec:level3}}

We begin with the following action\footnote{We want to emphasise that, in the absence of an explicit stringy-embedding of the effective action, we are assuming the existence of a large $N_c$ dual gauge theory. In the most conservative approach the reader may consider the phrase ``large $N_c$ gauge theory" as a placeholder for the dual description. This is perhaps not unreasonable since asymptotically AdS-spaces are potentially duals of some ``large $N_c$ gauge theories".}
\begin{eqnarray}\label{effaction2}
S = \frac{1}{8\pi G_N^{(4)}}\int d^4x\sqrt{- G}\left[\frac{1}{2}\left( R + 6 - \frac{1}{2}(\partial\phi)^2 \right) - \frac{h(\phi)}{4} F_{\mu\nu} F^{\mu\nu} \right] \ ,
\end{eqnarray}
which leads to the following equations of motion
\begin{eqnarray}
R_{\mu\nu}-\frac{1}{2} G_{\mu\nu} \left(R + 6 \right) & = & T_{\mu\nu} \ , \label{eqn1} \\
\nabla^2 \phi  - \frac{1}{2} F_{\mu\nu}F^{\mu\nu} \frac{\partial h(\phi)}{\partial \phi} & = &  0 \ , \label{eqn2} \\
\nabla_\nu \left(h(\phi) F^{\mu\nu} \right) & = &  0 \ , \label{eqn3}
\end{eqnarray}
where
\begin{eqnarray}
T_{\mu\nu} & = & T_{\mu\nu}^{\rm (scalar)} + T_{\mu\nu}^{\rm (Maxwell)} \ , \\
T_{\mu\nu}^{\rm (scalar)} & = & \frac{1}{2}(\partial_{\mu}\phi \partial_{\nu}\phi) - \frac{1}{4} G_{\mu\nu} (\partial\phi)^2 \ , \label{EMmax} \\
 T_{\mu\nu}^{\rm (Maxwell)} & = & h(\phi) F_{\mu\sigma} F^{\sigma}_{\nu} - \frac{1}{4} G_{\mu\nu} h(\phi) F_{\rho\sigma} F^{\rho\sigma} \ . \label{EMscalar}
\end{eqnarray}
Let us now specify our ansatz which consists of the metric field, the Maxwell field and the scalar field: $\{G, A, \phi\}$. We will assume translational invariance and we choose the ingoing Eddington-Finkelstein patch to represent our ansatz data:
\begin{align}
	ds^2 & = -\frac{2}{z^2} dv dz - g(z,v) dv^2 + f(z,v)^2 dx_i^2 \ , \label{metric0} \\
         \phi & = \phi(z,v) \ , \label{phi0} \\
         A & = A_v (z,v) dv \ . \label{A0}
\end{align}
We need two more sets of data: the boundary conditions and the initial conditions. The boundary conditions simply impose that the geometry is asymptotically locally AdS, which is represented by
\begin{eqnarray}
	g(z,v) & = & \frac{1}{z^2} \left(1 + \cO\left(z^2\right) \right) \qquad  {\rm as} \qquad  z\rightarrow 0 \ , \label{eq:bdyconditions1} \\
	f(z,v) & = & \frac{1}{z} \left (1+ \cO\left(z \right) \right) \qquad   {\rm as} \qquad z\rightarrow 0 \ , \qquad {\rm and} \label{eq:bdyconditions2} \\
	\phi(z,v) & = & \phi(v) + \cO(z)  \qquad   {\rm as} \quad z\rightarrow 0 \label{eq:bdyphi} \ , \\
	A_v (z, v) & = & {\rm const} + \cO(z) \ . \label{eq:bdyconditions4}
\end{eqnarray}
This choice fixes the gauge completely.\footnote{Demanding that $f(z,v)\sim \frac{1}{z} +\cO(1)$ fixes the gauge redundancy which remains after choosing Eddington-Finkelstein gauge $g_{z z}=0,\ g_{v x}=0,\ g_{z v }=1$. See {\it e.g.}~\cite{Bhattacharyya:2009uu}.}

Let us specify the initial condition now. Our initial state in the dual field theory corresponds to a thermal state with a non-vanishing chemical potential. This is represented by
\begin{eqnarray}
	\lim_{v \to -\infty} g(z, v) & = & \frac{1}{z^2} \left(1-M  z^3 + \frac{Q^2}{2} z^4 \right) \ , \label{eq:initial1} \\
	\lim_{v \to -\infty} f(z, v) & = & \frac{1}{z} \ , \label{eq:initial2} \\
	\lim_{v \to -\infty} \phi(z, v) & = & 0  \ ,  \label{eq:initial3} \\
	\lim_{v \to -\infty} A_v (z, v) & = & \mu_i + Q z \ . \label{eq:initial4}
\end{eqnarray}
\todo{}{\color{red} [AK: Represent the data in terms of causal infinite past, conformal boundary etc, or just the way above? ]} Our goal now is to find a solution of the system of equations in (\ref{eqn1})-(\ref{eqn3}) subject to the initial conditions in (\ref{eq:bdyconditions1})-(\ref{eq:bdyphi}) and with the asymptotically locally AdS boundary condition in (\ref{eq:initial1})-(\ref{eq:initial4}). We want to introduce dynamics in the system by exciting a time-dependent non-normalizable mode for the scalar field near the boundary, which can be represented by
\begin{equation}\label{eq:phi0}
	\phi_{\rm b} (v)=	\begin{cases}
		         &0 \qquad  \qquad \ v<0\\
			 &\epsilon\  \phi_1(v)  \qquad   0<v<\delta t\\
 &0 \qquad  \qquad \ v>\delta t \ , \\
 \end{cases}
 \end{equation}
 where $\phi_1(v)$ is now a function\footnote{Note that the symmetry and boundary behavior requirements allow the scalar $\phi$ to be an arbitrary function of time at the boundary.} of $\cO(1)$ and the dimensionless parameter $\epsilon \ll 1$ will eventually serve as the expansion parameter. To connect with the discussion in section \ref{sec:level2}, we note that: $\epsilon = \kappa_{\rm quench}^{1/2}$. \todo{}{\color{red}[AK: comment on the dimension of the field $\phi$ as well.]} Note that, the energy-momentum tensor in (\ref{EMmax}), (\ref{EMscalar}) evaluated on (\ref{eq:phi0}) is not of a null-dust-type, as considered in {\it i.e.}~\cite{Caceres:2012em}, and thus we will not encounter potential subtleties associated with violating null energy condition\cite{Caceres:2013dma}.

Before proceeding further, let us comment on the particular coordinate patch. To incorporate the dynamics, we have chosen the Eddington-Finkelstein coordinates, collectively denoted by $\cU_{\rm EF}$, which is well-defined everywhere. On the other hand, in order to read off the stress-energy tensor of the dual field theory, it is very useful to express all data in terms of the Fefferman-Graham patch, which we denote by $\cU_{\rm FG}$. We can define a map $\varphi : \cU_{\rm EF} \to \cU_{\rm FG}$, such that
\begin{eqnarray}
ds_{\rm EF}^2 & = & - \frac{2}{z^2} dz dv - g (z, v) dv^2 + f(z,v)^2 dx_i^2 \nonumber\\
                      & = & \frac{dr^2}{r^2} + \frac{\gamma_{ab}(x, r)}{r^2} dx^a dx^b = ds_{\rm FG}^2 \ , \quad {a,b} = 0 , \ldots 2
\end{eqnarray}
where
\begin{eqnarray}
\varphi \equiv \left \{z(t, r) , v(t, r) \right \} \ , \quad {\rm satisfying} \nonumber\\
\end{eqnarray}
\begin{eqnarray}
\dot{v} z' + v' \dot{z} + z^2 \dot{v} v' & = & 0 \ , \label{ef2fg1} \\
\frac{2}{z^2} v' z' + g(z, v) v'^2 & = & - \frac{1}{r^2} \ ,  \label{ef2fg2} \\
- r^2 \left( \frac{2}{z^2} \dot{v} \dot{z} + g(z,v) \dot{v}^2 \right) & = & \gamma_{00} \ , \label{ef2fg3}
\end{eqnarray}
where $' \equiv \partial_r$ and $ \dot{} \equiv \partial_t$. Near the boundary we have
\begin{eqnarray}
\gamma_{ab} (x, r) = \gamma_{ab} (x)  + r^3 \gamma_{ab}^{(3)} (x) + \ldots \ . \label{fgform}
\end{eqnarray}
After appropriate holographic renormalization, and using the GKPW recipe, the stress-energy tensor of the dual field theory is obtained to be \cite{de Haro:2000xn, Skenderis:2000in, Skenderis:2002wp}
\begin{eqnarray}
\langle T_{ab} \rangle = \frac{3}{16 \pi G_N^{(4)}} \gamma_{ab}^{(3)} \ . \label{ftst}
\end{eqnarray}
Evidently, once we obtain a solution in the $\cU_{\rm EF}$-patch, we can use the map $\varphi : \cU_{\rm EF} \to \cU_{\rm FG}$ by solving equations in (\ref{ef2fg1})-(\ref{ef2fg3}) and finally using (\ref{fgform}), (\ref{ftst}) we can read off the field theory stress-tensor of the corresponding state. In practice though, for the initial and the final states, the boundary energy-momentum tensor can be obtained by analyzing the thermodynamics of the corresponding state.

\subsection{Asymptotic, $z\rightarrow 0$,  expansion}\label{sec:z0expansion}

Let us first investigate the near  boundary behavior of the solution to \eqref{eqn1}-\eqref{eqn3} to ensure the asymptotically locally AdS criterion. As $z\rightarrow 0$, we introduce the formal expansion
\begin{eqnarray}
\nonumber
f(z,v)&=&\frac{1}{z}+\sum_{n=0}z^n\mk{f}_{n}(v) \ , \\
\nonumber
g(z,v)&=&\frac{1}{z^2}+ \sum_{n=0}z^n\mk{g}_{n}(v) \ , \\
\phi(z,v)&=&\phi_{\rm b}(v) +\sum_{n=1}z^n\mk{p}_{n}(v) \ , \\
\nonumber
A_v(z,v)&=&\sum_{n=0}z^n\mk{a}_{n}(v) \ ,
\end{eqnarray}
where $\phi_{\rm b}(v)$, as introduced in (\ref{eq:phi0}) before, denotes the source which we are turning on at the boundary and the set of functions $\{\mk{f}_{n}, \mk{g}_{n}, \mk{p}_{n}, \mk{a}_{n} \}$ can be systematically determined from the equations of motion at each order in $z$-expansion.

For illustrative purpose, we provide below explicit formulae up to $\cO(z^5)$-term
\begin{eqnarray} \label{eq:near-bdy-sol}
g(z,v) & = &  \frac{1}{z^2}\left[1-\frac{3}{4} \phi_{\rm b}'(v)^2z^2+M(v)z^3 + \left(\frac{c^2}{2 h(\phi_{\rm b})}-\frac{1}{24} \phi_{\rm b}'(v)^4+\frac{1}{2}L(v)\phi_{\rm b}'(v) \right)z^4+\mathcal{O}(z^5)\right] \ , \nonumber\\\label{eq:near-bdy-sol1} \\
f(z,v) & = & \frac{1}{z}\left[1-\frac{1}{8} \phi_{\rm b}'(v)^2z^2+\frac{1}{384} \left(-48L(v)\phi_{\rm b}'(v)+\phi_{\rm b}'(v)^4 \right)z^4 + \mathcal{O}(z^5)\right] \ , \label{eq:near-bdy-sol2} \\
\phi(z,v) & = & \phi_{\rm b}(v)+\phi_{\rm b}'(v) z + L(v) z^3 \nonumber\\
& - & \frac{1}{4} \left(\frac{c^2 \, \dot{h}(\phi_{\rm b})}{h(\phi_{\rm b})^2} + \phi_{\rm b}'(v) \left(M(v) + \phi_{\rm b}'(v) \phi_{\rm b}''(v)\right) + 4 L'(v)\right) z^4 + \mathcal{O}(z^5) \ ,  \label{eq:near-bdy-sol3} \\
A_v(z,v) & = & \mu(v) + \frac{c z}{h(\phi_{\rm b})} - \frac{c \, \dot{h}(\phi_{\rm b}) \phi_{\rm b}'(v)}{2 h(\phi_{\rm b})^2}z^2 \nonumber\\
& + & \frac{c \, \phi_{\rm b}'(v)^2 \left(h(\phi_{\rm b})^2 + 4 \dot{h}(\phi_{\rm b})^2 - 2 h(\phi_{\rm b}) \ddot{h} (\phi_{\rm b})\right)}{12 h(\phi_{\rm b})^3}z^3 + \mathcal{O}(z^4) \ ,  \label{eq:near-bdy-sol4}
\end{eqnarray}
where $c$ is a free parameter, $' \equiv \partial_v$ and $\dot{} \equiv \partial_\phi$. Furthermore, $L(v)$ and $M(v)$ are undetermined functions which satisfy
\begin{equation}
	M'(v) = - \frac{1}{8}  \phi _{\rm b}'(v) \left(3 \phi _{\rm b}'^3(v)-4 \phi_{\rm b}^{'''}(v) -12 L(v)\right) \ .
\end{equation}
The two time-dependent functions $M(v)$ and $L(v)$, which are not determined by the asymptotic expansion above, physically correspond to the mass of the black hole and the VEV of the marginal operator, respectively. Note that, the solutions in (\ref{eq:near-bdy-sol1})-(\ref{eq:near-bdy-sol4}) represent an asymptotic solution of the full geometry.

So far, we have imposed the asymptotically locally AdS condition in details. In order to carry out a perturbative treatment, we need to identify the ``correct" initial state around which it is meaningful to carry out a perturbation order by order. We will answer this question {\it a priori}: it turns out that one choice for which such a perturbative treatment works is to start from an AdS-RN geometry with a ``small" mass and a ``small" charge. Thus, instead of arbitrary $M$ and $Q$ in (\ref{eq:initial1}), we rewrite them as
\begin{eqnarray}
\{M, Q\} = \epsilon_1^2 \, \{m, q \} \ , {\quad} {\rm with} \quad \{m, q\} \sim \cO(1) \ ,
\end{eqnarray}
where $\epsilon_1 \ll 1$ is another small parameter. We can identify $ \epsilon_1 = \kappa_\mu^3$ to connect to the discussion in section \ref{sec:level2}. Clearly, $\epsilon$ and $\epsilon_1$ are hitherto independent parameters.

Thus, the initial state can be written as
\begin{alignat}{3}\label{eq:initial-pert}
	&g(z,v)=\frac{1}{z^2} \left(1 - m  \epsilon_1^2  z^3 + \frac{q^2 \epsilon_1^4}{2} z^4 \right) \qquad  \qquad & (v<0) \ , \\
	&f(z,v)=\frac{1}{z} & (v<0) \label{eq:initial-pert2} \ , \\
	& A_v(z,v)= \mu_i + q \epsilon_1^2 z  & (v<0) \  . \label{eq:initial-pert3}
\end{alignat}
Here $\mu_i$ can be fixed by demanding regularity of the gauge field at the event horizon.

\subsection{Expansion in amplitude of $\phi_{\rm b}(v)$}\label{sec:epsilonexpansion}

To solve the equations of motion, we now work with the following formal expansion
\begin{eqnarray}
f(z,v) & = & \sum_{n=0} \epsilon^n {f}_{n}(z,v) \ , \label{weakanstaz1} \\
g(z,v) & = & \sum_{n=0} \epsilon^n {g}_{n}(z,v) \ , \label{weakanstaz2} \\
\phi(z,v) & = & \sum_{n=0} \epsilon^n {\Phi}_{n}(z,v) \ , \label{weakanstaz3} \\
A_{v}(z,v) & = & \sum_{n=0} \epsilon^n A_{v_n} (z,v) \ , \label{weakanstaz4}
\end{eqnarray}
where the data $\left\{ f_n, g_n, \Phi_n, A_{v_n}\right\}$ are to be systematically determined from the equations of motion.

In order to perturbatively solve the equations of motion and motivated by convenience, we will treat $\epsilon_1$ as  a parameter of $\mathcal{O}(\epsilon)$. Thus,
\begin{eqnarray}
\epsilon_1 = k \, \epsilon \ , \quad {\rm with} \quad k \sim \cO(1) \label{defk} \ .
\end{eqnarray}
Note that the above relation is akin to the condition $\cO(\kappa_\mu) = \cO( \kappa_{\rm quench})$. It is clear that in this regime we regroup the dynamical contributions at various unrelated orders involving $\epsilon$ and $\epsilon_1$. Physically this corresponds to linearly combining processes which occur at {\it e.g.}~$\cO(\epsilon^2)$ and $\cO(\epsilon_1^2)$, and hence there is only one expansion parameter at the end.

We will be able to check  two independent limits of setting $\epsilon \to 0$ and $\epsilon_1 \to 0$ using  (\ref{defk}).
Solving the  dilaton equation of motion requires  $\dot{h}(0) = 0$. Similarly, from  Einstein field equation we obtain $h(0)=1$. To present the explicit solution up to, {\it e.g.}~$\cO(\epsilon^4)$, we first define the following functions
\begin{align}
		& C_2(v) = -\frac{1}{2} \int _{-\infty}^ v \phi_1'(x) \phi_1'''(x) dx  \label{eq:C2} \ , \\
		& C_4(v) = \frac{3}{8} \int_{-\infty}^v  \phi_1'(x) \left( - \phi_1'(x)^3 + \int_{-\infty}^x( B(y)-m k^2 \phi_1'(y))dy \right) dx \label{eq:C4} \ , \\
		& B(x) = \phi_1'(x) \left( C_2(x) + \phi_1'(x) \phi_1''(x) \right) \ , \\
		& P(v) = \frac{1}{4}\int_{-\infty}^v \left(- m k^2  \phi_1'(x)+ B(x) \right) dx \label{eq:P} \  , \\
		& a_4(v) = - q k^2  \int _{-\infty}^ v \phi_1 (x) \phi_1'(x) \ddot{h}(0) dx \label{eq:a4} \ .
	\end{align}
In order to properly account for the powers of $\epsilon$, let us write the $\epsilon$-dependence of $C_4(v), P(v)$ and $a_4(v)$ explicitly. From \eqref{eq:C4}, \eqref{eq:P} and \eqref{eq:a4} we have,
	\begin{align}
		&C_4(v)=c_4(v) + m k^2 \mk{c_4} (v)  \ ,  \\
		&P(v)=p(v)+m k^2 \mk{p}(v) \ , \\
		&a_4(v)= q k^2  \mk{a_4}(v) \ ,
	\end{align}
where now neither  ${c_4}, p $ nor $\mk{c_4} ,\mk{ p}, \mk{a_4}$ depend on $\epsilon$. With these definitions the solution can now be written as:\footnote{We are retaining the factors of $\epsilon$ and $\epsilon_1$ separately as a bookkeeping device.}
	\begin{align}
		& g(z,v) = \frac{1}{z^2} - m z \epsilon_1^2 + \frac{q^2 \epsilon_1^4}{2} z^2 -
		\left(z C_2(v) + \frac{3}{4} \phi_1'(v)^2 \right) \epsilon^2 \nonumber\\
		& \qquad  \quad +\left(z c_4(v) + \frac{z^2}{24} \left(12 p(v)\phi_1'(v)- \phi_1'(v)^4 \right) +
			\frac{z^3}{12} \left(3 p(v) \phi_1''(v)  - p'(v)\phi_1'(v) - C_2(v)  \phi_1'(v)^2 \right) \right) \epsilon^4 \nonumber\\
		&\qquad \quad  + m \left(z \mk{c_4}(v)+ \frac{z^2}{2}  \mk{p}(v)\phi_1'(v)+ \frac{ z^3}{12} \left( 3 \mk{p} (v)\phi_1''(v)- \mk{p}'(v)\phi_1'(v) - \phi_1'(v)^2 \right)  \right) \epsilon_1^2 \epsilon^2  +\mathcal{O}\left(\epsilon^6 \right) \ , \label{evo1} \\
		& f(z,v) = \frac{1}{z} - \frac{1}{8} z\phi_1'(v)^2 \epsilon^2 +
		\frac{z^3}{384} \left(
		\phi_1'(v) ^4-48  p(v)\phi_1'(v) \right) \epsilon^4 - \frac{z^3}{8}  \mk{p}(v)\phi_1'(v) \epsilon_1^2 \epsilon^2  +  \mathcal{O} \left(\epsilon^6 \right) \ ,  \label{evo2} \\
		& \Phi(z,v) = (\phi_1(v) + z\phi_1'(v)) \epsilon + z^3 p(v) \epsilon^3 + z^3 m  \mk{p}(v) \epsilon_1^2 \epsilon^2 +  \mathcal{O} \left(\epsilon^5 \right) \ ,   \label{evo3} \\
		& A(z,v) =  \mu(v)+ q z \epsilon_1^2  + \frac{q}{12}\left( 12 z \mk{a_4}(v) - 6  z^2 \phi_1(v)\phi_1'(v) \ddot{h}(0) + z^3 \phi_1'(v)^2 (1 -2 \ddot{h}(0)) \right) \epsilon_1^2 \epsilon^2 +  \mathcal{O} \left(\epsilon^4 \right) \ .  \label{evo4}
	\end{align}
We can now relate the boundary quantities  $\left \{M(v),\ L(v),\ c \right \}$ appearing in (\ref{eq:near-bdy-sol1}), (\ref{eq:near-bdy-sol3}) to the amplitude expansion:\todo{}{\color{red} [AK: I'm not sure yet where this part should go.]}

\begin{align}
& M(v) = - \left( m k^2 + C_2(v) \right) \epsilon^2
		+ \left( c_4(v) + m k^2 \mk{c_4}(v) \right) \epsilon^4 + \mathcal{O}(\epsilon^6)  \ ,  \\
& L(v)  = \left( p(v) + m k^2 \mk{p}(v) \right) \epsilon^3 +  \mathcal{O}(\epsilon^5) \ ,  \label{Lveq} \\
&\frac{c}{h(\phi_b(v))}= q k^2 \epsilon^2 + q k^2 \mk{a_4}(v) \epsilon^4 + \mathcal{O}(\epsilon^5) \ .
\end{align}
 \todo{}{\color{red}[AK: We need to comment on the map between Eddington-Finkelstein patch and the Fefferman-Graham patch and define what we mean by ``boundary quantities". Provide more details here.]}

Before going further, let us check a couple of trivial limits: (i) First, note that setting $\epsilon =0$ keeping $\epsilon_1 \not = 0$, we kill off the entire dynamics and get back the initial AdS-RN state in (\ref{eq:initial-pert})-(\ref{eq:initial-pert3}). In other words, this limit is equivalent to taking $v \to - \infty$. (ii) Secondly, if we set $\epsilon_1 = 0$ keeping $\epsilon \not = 0$, our initial state reduces to empty AdS. It is straightforward to check that this case reduces to the one with a vanishing chemical potential\cite{Bhattacharyya:2009uu}. However, the non-trivial dynamics remain. (iii) Third, if we set both $\epsilon_1 = 0 = \epsilon$, then we are left in the empty AdS-background with no dynamics. This is the most trivial limit of the solution described above. As alluded in (\ref{defk}),  we will consider $\epsilon_1 \not = 0$ and $\epsilon \not = 0$, with the constraint that $k = \epsilon_1 / \epsilon\sim \cO(1)$.

At late times, {\it i.e.}~$v \gg \delta t$, the final state is given by
\begin{align}\label{eq:finalfinal}
	& g(z,v) = \frac{1}{z^2} - z \left[ \epsilon^2 \left(m k^2 + \tilde{C}_2 \right) - \epsilon^4 \,  \wt{c}_4  \right] + z^2\frac{\epsilon^4k^4q^2}{2}+ \mathcal{O} \left(\epsilon^6 \right)  \ ,   \\
			& f(z,v) = \frac{1}{z}  + \mathcal{O} \left(\epsilon^6 \right) \ , \\
		& \Phi(z,v) =z^3 \,   \wt{p} \,  \epsilon^3 + \mathcal{O} \left(\epsilon^5 \right) \ ,\\
		&A(z,v)= \mu_f + q \, z\,   k^2 \epsilon^2
	\ . \label{eq:finalfinal4}
	\end{align}
	Here the tilde denotes  the function  evaluated at any $v  \gg \delta t$. Since $\phi_1$ is of compact support, this is equal to the value of the function  at infinity,  $\wt{C}_2 = C_2(\infty) = -\frac{1}{2} \int _{-\infty}^ \infty \phi_1'(x) \phi_1'''(x) dx$,\ {\it etc}. Note that to fourth order,  the mass of the black hole has increased by an amount $\wt{C}_2\, \epsilon^2 +  \wt{c}_4 \, \epsilon^4$ as compared to the initial state, similarly to \cite{Bhattacharyya:2009uu}. Exploring the solution at higher order we find that at sixth order  $g(z,v)$ has a dependence on $\ddot{h}(0)$ that survives at late time and gives a subleading contribution to the mass of the final black hole.  Note also that  the gauge field at late times differs from the original gauge field only by a shift in the chemical potential. This is consistent with our expectations since the charge of the black hole remains the same. However,  the position of the horizon changes and thus --- as we will see in section \ref{sec:level4} --- the chemical potential also changes due to the boundary condition on the gauge field. We will analyze the resulting thermodynamics in Section \ref{sec:level4}.

\subsection{\label{sec:level5}Analytic structure and regime of validity}

Let us briefly investigate  the analytic structure and regime of validity  of the perturbative solution for $v>\delta t$. We can systematically find the solution to arbitrary  order in  $\epsilon$. Then,  following a similar approach as in \cite{Bhattacharyya:2009uu}, we can inductively show that, among the functions that appear in \eqref{weakanstaz1}-\eqref{weakanstaz4},
\begin{equation}
g_{2n+1} = 0 \ , \quad f_{2n +1} = 0 \ , \quad \phi_{2n} = 0 \ , \quad A_{v_{2n+1}} = 0 \ , \quad \forall n \in \mathbb{Z}_+ \ .
\end{equation}
Thus the non-trivial information about the dynamics is contained in the set of functions: $ {\mathbb G} \equiv \left\{{\mathbb G}_n \right\} = \left \{g_{2n}, f_{2n}, \phi_{2n+1}, A_{v_n} \right \}$ and they take the following general form

\begin{align}\label{eq:epsilonexpansion}
		& \phi_{2n+1}(z,v) = \sum_{k=0}^{2n-2} z^{2n+1-k} \phi^k_{2n+1}(v) \ , \qquad n\ge2 \nonumber\\
		& f_{2n}(z,v) = \frac{1}{z} \sum_{k=0}^{2n-4} z^{2n-k} f^k_{2n}(v) \ , \qquad n\ge3 \\
		& g_{2n}(z,v) = z  C_{2n}(v)  + \frac{1}{z} \sum_{k=0}^{2n - 3} z^{2n-k} g_{2n}^k(v) \ , \qquad  n\ge3 \nonumber\\
		& A_{v_{2n}}(z,v) = \mu_{2n}(v) + \sum_{k=0}^{2n-2} z^{2n-1-k} A_{v_{2n}}^k(v) \ .
		\qquad n\ge 3\nonumber
	\end{align}
In general $\mathbb{G}$ is a function of $v$ and a functional of $\phi_1(v)$ and its derivatives: $\mathbb{G} = \mathbb{G} \left(v, \phi_1, \phi_1' \, , \ldots\right)$.
Note  that even powers of $\epsilon$ are absent in $\phi(z,v)$ and odd powers are absent  in $g(z,v) $, $f(z,v)$ and $A_v(z,v)$. Likewise, the odd derivatives of the coupling function, $h(\phi)$, evaluated at $\phi=0$ should vanish, $0=\dot{h}(0)=\dddot{h}(0)=h^{(5)}(0)... = 0$.\footnote{As stated previously, the equations of motion demand that  $h(0)= \dot{h}(0)=0$. The requirement that the odd higher derivatives  vanish ( $\dddot{h}(0)=h^{(5)}(0)=h^{(2j+1)}...=0$) comes from assuming a series of the form \ref{eq:epsilonexpansion} for the dilaton and the gauge field. }

For $v>\delta t$ the functions $\tilde{ {\mathbb G}}= \left \{g_{2n}, f_{2n}, \phi_{2n+1} \right \}$ consists of polynomials in $v$ of a degree that grows with $n$ \footnote{The late time behavior of $A_v$ is different; all terms of order $\epsilon^4 $  or higher  go to zero at late times. Thus, at  $v>\delta t$ we recover the original gauge field and $A_v$ does not enter in the analysis of the late time validity of the solution.}. In particular $\phi_{2n+1}$ is at most of degree $(n+k-1)$, $f_{2n}$ is at most of degree $(n+k-3)$ and $g_{2n}^k$ of degree $(n+k-4)$. Thus we have
\begin{eqnarray}
	{\rm max} \, \left  \{ {\rm deg} \left[ {\tilde{\mathbb G}}_n\right] \right \} = (n+k-1) \ .
\end{eqnarray}
The fact that the functions $\tilde{{\mathbb G}}_n$ are polynomials in $v$ whose degree grows with $n$  implies that the series expansion will break down at late times. In order to characterize the regime of validity lets focus on $\phi(z,v)$, which has the maximum degree in $v$:
	\begin{equation} \label{eq:genexpansion}
		\phi(z,v) = \sum_{n,k} \epsilon^{2n+1} \phi_{2n+1}^k \, z^{2n+1-k} \ .
	\end{equation}
It can further be checked that for late times
\begin{eqnarray}
\phi_{2n+1}^k \sim\frac{v^{n+k-1}}{\delta t^{3n}} \ .
\end{eqnarray}
It would suffice for our purposes, if the perturbative solution is valid up to the event-horizon of the geometry. Recall that the event-horizon is given by, up to the leading order in $\epsilon$,
\begin{equation}
		z_{\rm H} \sim \frac{1}{\epsilon^{2/3}(k^2 m + \tilde{C}_2) ^{1/3}} \ .
\end{equation}
Also recall that $k$ and $m$ are  $\mathcal{O}(1)$ numbers while $ \tilde{C}_2\sim \frac{1}{\delta t ^3 }$. Thus we get: $z_{\rm H} \sim\frac{\delta t}{ \epsilon^{2/3}}$. Therefore,  close to the horizon and for late times, the term with labels $n, k$ in  \eqref{eq:genexpansion} will scale approximately as $(\epsilon^{2/3} \frac{v}{\delta t})^{n-1+k} \epsilon $. This implies that if $ \epsilon^{2/3} \frac{v}{\delta t} \ll 1$ the small values of $\{n, k\}$ dominate the series and larger values are subleading. However if $ \epsilon^{2/3} \frac{v}{\delta t} \gg 1$ it is the larger values of $\{n, k\}$ that dominate and the perturbation series breaks down. Thus, our series solution is good only up to times $v\sim \mathcal{O} \left( \frac{\delta t}{\epsilon^{2/3}} \right)$ which  to leading order in $\epsilon$ corresponds to $1 / \left( \Delta E \right)^{1/3}$  (see eq.~(\ref{temp2})).

\setcounter{equation}{0}
\section{Thermodynamics of the states\label{sec:level4}}

We will now briefly discuss the initial and the final states which are interpolated by the evolution described in (\ref{evo1})-(\ref{evo4}). Both our initial and the final states are characterized by a non-vanishing temperature and a chemical potential, both of which have dimension ${\rm length}^{-1}$. It is also accompanied by the VEV of the marginal operator, denoted by $\langle \cO_\phi\rangle \sim {\rm length}^{-3}$, which is being quenched {\it via} the dynamics. Thus the state is specified by the following data:
\begin{eqnarray}
{\rm QFT} \ni \left\{ \mu , T ,  \langle \cO_\phi \rangle \right\}  \quad \Longleftrightarrow  \quad  \left\{ G, A, \phi \right\} \in {\rm Gravity} \ .
\end{eqnarray}
Furthermore, note that from the full solution in (\ref{evo4}) it is clear that the normalizable mode of the gauge field does not have any dynamics associated and thus
\begin{eqnarray}
\lim_{z \to 0} \partial_v F_{z v} = \partial_v \left( q k^2 \epsilon^2\right) = 0  \ ,
\end{eqnarray}
which implies that the charge density in the dual field theory is kept fixed. Thus, we are considering the dynamical evolution in a ``canonical ensemble". We will now discuss the initial and the final states in more detail.

The regularized on-shell Euclidean action corresponds to the Gibbs free energy, which characterizes the grand-canonical ensemble. If we denote the Gibbs free energy by $W$, then the Helmholtz free energy, which we denote by $F$, characterizes the canonical ensemble and is obtained by a Legendre transformation of the Gibbs potential
\begin{eqnarray}
F = W - \mu \, Q \ ,
\end{eqnarray}
where $\mu$ and $Q$ are respectively the chemical potential and the charge density.

\subsection{Initial state}

Here we will reinstate $\epsilon_1$ where they originally appear. According to (\ref{eq:initial-pert})-(\ref{eq:initial-pert3}), in the limit $v\to-\infty$, we get:
\bea
f(z) & = & \frac{1}{z} \,,\\
g(z) & = & \frac{1}{z^2}\left(1-\epsilon_1^2 m z^3+\frac{\epsilon_1^4 q^2 z^4}{2}\right) \,,\\
A_v(z) & = & \mu_i+\epsilon_1^2 q z \,,\\
\phi(z) & = & 0 \,.
\eea
With the scalar field turned off, the geometry reduces to the usual AdS-RN black hole. For the above solution, the corresponding mass $M$ and charge $Q$ are given by
\be\label{masscharge}
M = \epsilon_1^2 m\quad\text{and}\quad Q^2=\epsilon_1^4q^2 \,.
\ee
Alternatively, the mass can also be given as
\be
M=\frac{1}{z_{\rm H}^3}+\frac{Q^2}{2}z_{\rm H} \,,
\ee
where $z_{\rm H}$ is the location of the event horizon, given by the smallest real positive positive root of the algebraic equation $g(z)=0$. For small $\epsilon_1$ we obtain,\footnote{For concreteness, we will evaluate all physical quantities in two leading order terms in $\epsilon_1$.}
\be
z_{\rm H} = \frac{1}{\epsilon_1 ^{2/3}m^{1/3}}\left(1+\frac{\epsilon_1 ^{4/3}q^2}{6 m^{4/3}}+\mathcal{O}(\epsilon_1 ^{8/3})\right)\,.
\ee
On the other hand, the one-form $A_v$ must be regular at the horizon such that $|| A ||$ remains finite at the bifurcation point of the Kruskal-extended patch. This imposes a constraint, relating the chemical potential to the charge and the mass:
\be\label{chemical}
\lim_{z \to z_{\rm H}} A = 0 \quad \implies \quad \mu_i = -\epsilon_1^2 q z_{\rm H} = -\frac{\epsilon_1^{4/3}q}{m^{1/3}}\left(1+\frac{\epsilon_1 ^{4/3}q^2}{6 m^{4/3}}+\mathcal{O}(\epsilon_1 ^{8/3})\right)\,.
\ee
The subscript $i$ in all subsequent physical quantities will stand for the initial state. To calculate the Hawking temperature of the initial state $T_i$, we first perform a Wick rotation obtained as usual by the replacement $t\to i \tau$. Since the Euclidean time direction shrinks to zero size at $z = z_{\rm H}$, we must require that $\tau$ be periodically identified with appropriate period $\beta_i$, \emph{i.e.} $\tau\sim \tau + \beta_i$. A simple calculation shows that
\be \label{tempini}
T_i = -\frac{1}{4\pi}\frac{d}{dz}g(z)\bigg|_{z_{\rm H}} = \frac{3}{4\pi z_{\rm H}}\left(1-\frac{1}{6}Q^2 z_{\rm H}^4\right) = \frac{3 \epsilon_1 ^{2/3}m^{1/3}}{4 \pi }\left(1-\frac{\epsilon_1 ^{4/3}q^2 }{3m^{4/3}}+\mathcal{O}(\epsilon_1 ^{8/3})\right)\,,
\ee
or equivalently,
\be\label{temp}
\beta_i\equiv\frac{1}{T_i}=\frac{4 \pi }{3 \epsilon_1 ^{2/3}m^{1/3}}\left(1+\frac{\epsilon_1 ^{4/3}q^2 }{3m^{4/3}}+\mathcal{O}(\epsilon_1 ^{8/3})\right)\,.
\ee
It is convenient to invert the relations $\mu_i(m,q)$ and $\beta_i(m,q)$ given in (\ref{chemical})-(\ref{temp}) to obtain $m(\mu_i,\beta_i)$ and $q(\mu_i,\beta_i)$. However, we have to proceed with some care given that $\mu_i\sim\mathcal{O} (\epsilon_1^{4/3} )$ whereas $\beta_i\sim\mathcal{O} (\epsilon_1^{-2/3})$. First we define rescaled quantities $\tilde{\mu}_i=\epsilon_1^{-4/3}\mu_i$ and $\tilde{\beta}_i=\epsilon_1^{2/3}\beta_i$ such that they both are of order $\mathcal{O} \left(\epsilon_1^0\right )$. Then, an expansion in $\epsilon_1$ is reliable and the inversions can be found perturbatively. To our order of approximation we find,
\be\label{minitial}
m=\frac{64 \pi ^3}{27 \tilde{\beta}_i^3}\left(1+\frac{9\epsilon_1^{4/3}\tilde{\beta}_i^2\tilde{\mu}_i^2}{16 \pi^2}+\mathcal{O}(\epsilon_1^{8/3})\right)\approx\frac{64 \pi^3 }{27 \epsilon_1 ^2\beta^3_i}\left(1+\frac{9\beta_i^2 \mu_i^2}{16 \pi ^2}\right)\,,
\ee
and
\be
q=-\frac{4 \pi\tilde{\mu}_i}{3 \tilde{\beta}_i}\left(1+\frac{3\epsilon_1 ^{4/3} \tilde{\beta}_i^2 \tilde{\mu}_i^2}{32 \pi^2}+\mathcal{O}(\epsilon_1^{8/3})\right)\approx-\frac{4\pi \mu_i}{3 \epsilon_1 ^2 \beta_i}\left(1+\frac{3 \beta_i^2 \mu_i^2}{32 \pi ^2}\right)\,.
\ee
To study the thermodynamics of these solutions, we first evaluate the Euclidean action $I$ on-shell which defines the grand canonical (Gibbs) potential $W=I/\beta_i$ (see for instance \cite{Chamblin:1999tk}). With our conventions, the full Euclidean action is given by analytically continuing (\ref{effaction2}). Moreover, when the space is asymptotically AdS the Gibbons-Hawking boundary term gives a vanishing contribution.

As usual, the action diverges upon integration, given that the volume of any asymptotically AdS geometry goes to infinity near the boundary. These divergences can be eliminated by subtracting the pure AdS contribution, obtaining
\bea
I & = & -\frac{\mathrm{Vol}(\mathbb{R}^2)\beta_i}{\k^2}\left[\int_0^{z_{\rm H}} \frac{\epsilon_1 ^4q^2}{2}dz-\int_{z_{\rm H}}^{\infty}\frac{3}{z^4}dz\right] \,, \\
& = & -\frac{\mathrm{Vol}(\mathbb{R}^2) \beta_i   m\epsilon_1 ^2}{\kappa ^2}\left(1+\frac{\epsilon_1 ^{8/3}q^4}{4 m^{8/3}}+\mathcal{O}\left (\epsilon_1^{4}\right )\right) \,.
\eea
This may be rewritten entirely in terms of $\beta_i$ and $\mu_i$ as
\bea
I&=&-\frac{64 \pi ^3 \mathrm{Vol}(\mathbb{R}^2) \epsilon_1 ^{4/3}}{27 \kappa ^2 \tilde{\beta }_i^2}\left(1+\frac{9 \epsilon_1 ^{4/3}  \tilde{\beta }_i^2 \tilde{\mu }_i^2}{16\pi ^2}+\mathcal{O}(\epsilon_1^{8/3})\right)\,,\\
&\approx&-\frac{64 \pi^3 \mathrm{Vol}(\mathbb{R}^2) }{27  \kappa ^2\beta_i ^2}\left(1+\frac{9 \beta_i ^2 \mu_i ^2}{16 \pi^2}\right)\,.\label{grand}
\eea

The grand canonical potential is given by $W=E_i-T_iS_i-\mu_i Q_i.$ Using the expression given in (\ref{grand}), we may compute the state variables of the
system. At leading order we get:\footnote{In RN black holes $S\to$ constant as $T\to0$, indicating the degeneracy of the ground state. This can be seen at our order of approximation from (\ref{entropini}) and (\ref{minitial}). A brief computation leads to $S_i\sim\mu_i^2$ as $T_i\to0$.}
\bea
E_i&=& \left(\frac{\partial I}{\partial\beta_i}\right)_{\mu_i}
-\frac{\mu_i}{\beta_i}\left(\frac{\partial I}{\partial\mu_i}\right)_{\beta_i}
\approx\frac{128 \pi ^3 \mathrm{Vol}(\mathbb{R}^2)}{27 \kappa ^2\beta_i ^3} \left(1+\frac{9 \beta_i ^2 \mu_i ^2}{16 \pi ^2}\right)\approx\frac{2 \mathrm{Vol}(\mathbb{R}^2) \epsilon_1 ^2 m}{\kappa ^2}\,, \label{enerini} \\
S_i&=&
\beta_i\left(\frac{\partial
I}{\partial\beta_i}\right)_{\mu_i}-I\approx\frac{64 \pi ^3 \mathrm{Vol}(\mathbb{R}^2)}{9 \kappa ^2\beta_i ^2} \left(1+\frac{3 \beta_i ^2 \mu_i ^2}{16 \pi ^2}\right)\approx\frac{4 \pi  \mathrm{Vol}(\mathbb{R}^2) \epsilon_1 ^{4/3}m^{2/3}}{\kappa ^2}\,, \label{entropini} \\
Q_i&=&-\frac{1}{\beta_i}\left(\frac{\partial
I}{\partial\mu_i}\right)_{\beta_i}\approx\frac{8 \pi  \mathrm{Vol}(\mathbb{R}^2) \mu_i}{3 \kappa ^2\beta_i }\approx-\frac{2 \mathrm{Vol}(\mathbb{R}^2) \epsilon_1 ^2q}{\kappa ^2}
\,.
\eea
Together, they indeed satisfy the first law of thermodynamics, $dE_i = T_i dS_i + \mu_i dQ_i$. Moreover, the free energy $W=I/\beta_i$ is always negative, indicating stability of the solutions. Notice that we could have alternatively worked in the canonical ensemble, which is characterized by the Helmholtz free energy $F=E-ST$. A brief computation shows that, at the same order of approximation,
\be
F\approx-\frac{64 \pi^3 \mathrm{Vol}(\mathbb{R}^2) }{27  \kappa ^2\beta_i ^3}\left(1-\frac{9 \beta_i ^2 \mu_i ^2}{16 \pi^2}\right)\,.
\ee

Let us now comment on the identification of the field theory quantity corresponding to the putative small parameter $\epsilon_1$. From (\ref{tempini}), (\ref{chemical}) and (\ref{enerini}), it is clear that parametrically $T_i / E_i \sim \cO(1)$. However,
\begin{eqnarray}
\frac{\mu_i}{T_i} = - \left( \frac{4\pi}{3} \right) \frac{q}{m^{2/3}} \epsilon_1^{2/3} \quad \implies \quad \epsilon_1 = \left( \frac{\mu_i}{T_i} \right)^{3/2} \cO(1) \ .
\end{eqnarray}
Thus, we are considering an initial state in which the chemical potential is small compared to the temperature. It also becomes clear that, in obtaining the perturbative solution outlined in (\ref{evo1})-(\ref{evo4}), we have set the expansion parameter $\epsilon \sim \left( \frac{\mu_i}{T_i} \right)^{3/2}$. By analyzing the final state, we will now observe that $\epsilon$ corresponds to an otherwise independent parameter in the dual field theory.

\subsection{Final state}

Let us now consider the final state. For $v\to\infty$ (in practice for $v \gg \delta t$) the bulk solution in (\ref{eq:finalfinal})-(\ref{eq:finalfinal4}) can be written as:
\bea
f(z)&=&\frac{1}{z}+\mathcal{O}(\epsilon^6)\,,\\
g(z)&=&\frac{1}{z^2}\left(1 - \epsilon^2 \left( k^2 m  +  \wt{C}_2 \right) z^3 + \epsilon^4 \left(\wt{c}_4 z^3 + \frac{k^4q^2}{2} z^4 \right)  \right) +  \mathcal{O} \left(\epsilon^6 \right) \,,\\
A_v(z) & = & \mu_f + \epsilon^2 k^2 q z   \,,\\
\phi(z)&=&z^3 \wt{p} \epsilon^3 +  \mathcal{O} \left(\epsilon^5 \right) \,,
\eea
where $k = (\epsilon_1/\epsilon)$ and all the tildes denote the corresponding functions evaluated at $v\to \infty$, which are of order $\mathcal{O}\left(\epsilon^0 \right)$.

The mass and charge of the final background are now
\be\label{masscharge2}
M_f = \epsilon^2(k^2 m+\wt{C}_2) - \epsilon^4\wt{c}_4 \quad\text{and} \quad Q_f^2 = \epsilon^4k^4q^2 = Q_i^2 \,,
\ee
which gives
\begin{eqnarray}
\Delta M = M_f - M_i = \epsilon^2 \wt{C}_2 - \epsilon^4 \wt{c}_4 \ .
\end{eqnarray}
The new horizon lies at
\be
z_{\rm H}^{(f)} = \frac{1}{\epsilon ^{2/3}m_2^{1/3}}\left(1+\frac{\epsilon ^{4/3}k^4q^2}{6 m_2^{4/3}} + \mathcal{O} \left(\epsilon ^{2} \right) \right) \,.
\ee
For convenience, we have defined $m_2=k^2 m+\wt{C}_2$. Regularity of the gauge field at the horizon now yields
\be\label{chemical2}
\lim_{z \to z_{\rm H}^{(f)}} A = 0 \quad \implies \quad \mu_f = -\frac{\epsilon^{4/3}k^2q}{m_2^{1/3}}\left(1+\frac{\epsilon^{4/3}k^4q^2}{6 m_2^{4/3}}+\mathcal{O} \left(\epsilon^{2} \right) \right) \,.
\ee
On the other hand, the inverse temperature is given by
\be\label{temp2}
\beta_f\equiv\frac{1}{T_f}=\frac{4 \pi }{3 \epsilon ^{2/3}m_2^{1/3}}\left(1+\frac{\epsilon ^{4/3}k^4q^2 }{3m_2^{4/3}}+\mathcal{O} \left(\epsilon ^{8/3} \right) \right)\,.
\ee
Defining $\tilde{\mu}_f=\epsilon^{-4/3}\mu_f$ and $\tilde{\beta}_f=\epsilon^{2/3}\beta_f$, we can invert (\ref{chemical2})-(\ref{temp2}) as follows
\bea
m_2 & = & \frac{64 \pi ^3}{27 \tilde{\beta}_f^3}\left(1+\frac{9\epsilon^{4/3}\tilde{\beta}_f^2\tilde{\mu}_f^2}{16 \pi^2}+\mathcal{O} \left(\epsilon^{8/3}\right ) \right) \approx \frac{64 \pi^3 }{27 \epsilon ^2\beta_f^3}\left(1+\frac{9\beta_f^2 \mu_f^2}{16 \pi ^2}\right) \,,
\eea
and
\be
q=-\frac{4 \pi\tilde{\mu}_f}{3 k^2\tilde{\beta}_f}\left(1+\frac{3\epsilon^{4/3} \tilde{\beta}_f^2 \tilde{\mu}_f^2}{32 \pi^2}+\mathcal{O} \left(\epsilon^{8/3}\right) \right) \approx -\frac{4\pi \mu_f}{3 \epsilon^2 k^2 \beta_f}\left(1+\frac{3 \beta_f^2 \mu_f^2}{32 \pi ^2}\right)\,.
\ee

After subtracting the AdS contribution, the Euclidean on-shell action evaluates to:
\bea
I&=&-\frac{ \mathrm{Vol}(\mathbb{R}^2) \beta  m_2 \epsilon ^2}{\kappa ^2}\left(1-\frac{3 \epsilon ^2(k^2m\wt{\mk{p}} + \wt{p})^2 }{4 m_2^2}+\mathcal{O} \left(\epsilon^{8/3}\right) \right) \,,\nonumber\\
&\approx& -\frac{64 \pi^3 \mathrm{Vol}(\mathbb{R}^2) }{27  \kappa ^2\beta_f ^2}\left(1+\frac{9 \beta_f ^2 \mu_f ^2}{16 \pi^2}\right)\,.
\eea

Finally, we can compute the state variables for the final state. At leading order we get:
\bea
E_f&=& \left(\frac{\partial I}{\partial\beta_f}\right)_{\mu_f}
-\frac{\mu_f}{\beta_f}\left(\frac{\partial I}{\partial\mu_f}\right)_{\beta_f}
\approx\frac{2 \mathrm{Vol}(\mathbb{R}^2) \epsilon ^2 m_2}{\kappa ^2}\,, \nonumber\\
S_f&=&
\beta_f\left(\frac{\partial
I}{\partial\beta_f}\right)_{\mu_f}-I\approx\frac{4 \pi  \mathrm{Vol}(\mathbb{R}^2) \epsilon ^{4/3}m_2^{2/3}}{\kappa ^2}\,,\nonumber\\
Q_f&=&-\frac{1}{\beta_f}\left(\frac{\partial
I}{\partial\mu_f}\right)_{\beta_f}\approx-\frac{2 \mathrm{Vol}(\mathbb{R}^2) \epsilon ^2k^2q}{\kappa ^2}
\,.
\eea
Again, these quantities satisfy the first law of thermodynamics, $dE_f = T_f dS_f + \mu_f dQ_f$. Note that, upon integration by parts, we get that $\wt{C}_2=\frac{1}{2}\int _{-\infty}^ \infty \phi_1''(x)^2 dx>0$ so we always have $m_2>m$. Both the energy and entropy increase but the charge remains the same, as expected. Moreover, for this final state, the Helmholtz free energy is given by
\be
F\approx-\frac{64 \pi^3 \mathrm{Vol}(\mathbb{R}^2) }{27  \kappa ^2\beta_f ^3}\left(1-\frac{9 \beta_f ^2 \mu_f ^2}{16 \pi^2}\right)\,.
\ee

Before concluding this section, let us comment on the parameter $\epsilon$. It is straightforward to check that
\begin{eqnarray}
\wt{C}_2 & \sim & \frac{1}{\left( \delta t \right)^3} \ , \quad {\rm and} \nonumber \\
\Delta E & = &  \left(E_f - E_i \right)  \sim  \frac{\epsilon^2}{\left( \delta t \right)^3}  \quad \implies \quad \epsilon \sim \left( \delta t \right)^{3/2} \left(\Delta E\right)^{1/2} \ . \label{epsmean}
\end{eqnarray}
Note that, this scaling is in keeping with the scaling behaviour obtained in \cite{Buchel:2013gba,Das:2014jna,Das:2014hqa}, which follows from dimensional analysis in our case. Evidently, (\ref{epsmean}) provides us with a natural meaning for the expansion parameter in gravity purely in terms of the field theory data. What is more, we observe that the perturbative solution is consistent as long as we impose
\begin{eqnarray}
\epsilon_1 \sim \epsilon \quad \implies \quad   \left( \frac{\mu_i}{T_i} \right)^{3} \sim  \left( \delta t \right)^{3} \left(\Delta E\right) \ .
\end{eqnarray}
%

\setcounter{equation}{0}
\section{\label{sec:level6}Thermalization time}

Given the background that we obtained in the previous section, we will now explore how the thermalization time behaves with relevant parameters for the system. We will measure thermalization time by measuring non-local observables, specially entanglement entropy and define our thermalization time according to the behavior of this non-local probe. Via the AdS/CFT correspondence, this exercise amounts to computing an extremal area surface in the bulk geometry. For this purpose, we will only require the solution for the metric (\ref{metric0}), which is given explicitly in (\ref{eq:finalfinal})-(\ref{eq:finalfinal4}). The specific form of $f(z,v)$ and $g(z,v)$ will not be important for now.

The metric (\ref{metric0}) is asymptotically AdS. To make this manifest, let us rescale the metric functions such that
\be\label{metric2}
ds^2=\frac{1}{z^2}\left[- 2dvdz - \tilde{g}(z,v)dv^2 + \tilde{f}^2(z,v)\left(dx^2+dy^2\right)\right] \ ,
\ee
where we have defined $\tilde{f}(z,v)\equiv z f(z,v)$ and $\tilde{g}(z,v)\equiv z^2g(z,v)$. These new functions satisfy that $\tilde{f}\to1$ and $\tilde{g}\to1$ as $z\to0$. Also, the boundary time coordinate is related to the Eddington-Finkelstein coordinates in (\ref{metric2}) through
\footnote{Here we neglect the redshift effect indicated in \cite{Caceres:2012px}. In the thin-shell limit for $v_0\rightarrow 0$, the redshift
factor will reduce to one for the regime outside the shell. Note that the $t_{\rm crit}$ introduced in (\ref{criticalt}) will be slightly longer when incorporating the redshift effect.}
\be\label{efcoords}
dv = dt - \frac{dz}{\tilde{g}(z,v)} \ .
\ee
This relation will be important below, for the computation of the thermalization time.

\subsection{Entanglement entropy}

We will compute entanglement entropy in this background for a particular shape: namely a ``rectangular strip" which can be parametrized by $\{x \in (- \ell/2, \ell/2) \} \cup \{y \in (- \ell_\perp/2, \ell_\perp/2)\}$. In a quantum system, entanglement (or geometric) entropy of a region $A$ with its complement $B$ is defined using the reduced density matrix {\it a la} von Neuman: $S_A = - {\rm tr}_A \, \rho_A \log \rho_A $, where $\rho_A$ is obtained by tracing over the degrees of freedom in $B$, $\rho_A = {\rm tr}_{B} \, \rho$.

In AdS$_{d+1}$/CFT$_d$, the covariant prescription to compute entanglement entropy for a given region $A$ was proposed by Hubeny, Rangamani and Takayanagi in
\cite{Hubeny:2007xt}, generalizing the originally proposed Ryu-Takayanagi formula \cite{Ryu:2006bv}:
\be \label{rt}
S_A = \frac{1}{4 G_N^{(d+1)}} {\rm ext} \left[ {\rm Area} \left(\gamma_A \right)\right] \ ,
\ee
where $G_N$ is the bulk Newton's constant, and $\gamma_A$ is the $(d-1)$-dimensional extremal-area surface such that $\partial \gamma_A = \partial A$.\footnote{We note that at this point the evidence in favor of the Ryu-Takayanagi proposal is much stronger than the covariant HRT proposal. However, till this issue is settled, we are assuming that HRT proposal is the correct prescription in our case.} Before delving into computation, let us flesh out the approximation in which we will be working for the rest of this article.

The perturbative solution that we have obtained in (\ref{evo1})-(\ref{evo4}) can be numerically subtle to handle, specially since we have a small parameter $\epsilon$. Instead, we make use of the small parameter in the following manner: The metric data can be summarized as follows:
\begin{eqnarray}
G_{\mu\nu} = G_{\mu\nu}^{(0)} + \epsilon^2 G_{\mu\nu}^{(1)} + \epsilon^4 G_{\mu\nu}^{(2)} + \ldots \ ,
\end{eqnarray}
Subsequently the geodesics and the entanglement entropy can also be determined by a similar expansion
\begin{eqnarray}
\gamma_A & = & \gamma_A ^{(0) } + \epsilon^2 \gamma_A^{(1)} + \epsilon^4 \gamma_A^{(2)} + \ldots \  , \\
S_A  & = & S_A ^{(0) } + \epsilon^2 S_A^{(1)} + \epsilon^4 S_A^{(2)} + \ldots \  .
\end{eqnarray}
It can be checked that upon using the equations of motion at the first non-trivial order in $\epsilon$, which in this case happens to be at $\cO(\epsilon^2)$, $\gamma_A^{(0)}$ determines $S_A^{(1)}$ completely \cite{Nozaki:2013wia,Pedraza:2014moa}. Thus, if we limit ourselves to results at the first non-trivial order in $\epsilon$, the task of determining geodesics simplifies significantly. On the other hand, it is clear from the metric in {\it e.g.}~(\ref{evo1}), the effect of the charge enters at $\cO(\epsilon^4)$, at which order there is no such simplification. In this article, for simplicity, we will keep our discussions limited to $\cO(\epsilon^2)$. We therefore emphasize that, our numerical results should be taken as {\it indications} of a certain physics rather than an {\it observation}. On the other hand, it is possible to go beyond $\cO(\epsilon^2)$ systematically, which we leave for future work.

Also note that, as we are measuring the {\it response} of the system at the first non-trivial order in $\epsilon$, so we will measure the {\it external dial}. In our case, the {\it external dial} is a combination of {\it e.g.}~$(\mu/T)\sim\cO(\epsilon^{2/3})$. Thus, effectively we are using a different precision for measuring the response-observables as compared to the parameters of the system. In a very precise sense, the subsequent dynamics that we analyze is primarily determined by the temperature-scale of the system.

Now let us discuss the details. We will now compute $S_A$ in the limit $\ell_\perp\to \infty$, so that the construction becomes invariant under translations in $y$. We can parameterize the extremal surfaces with functions $z(x)$, $v(x)$ subject to the boundary conditions
\be
z(\pm\ell/2)=z_0\quad\text{and}\quad v(\pm\ell/2)=t \ ,
\ee
 where $z_0$ is the usual UV cut-off needed to regularize the on-shell acion. These boundary conditions impose that the boundary of $\gamma_A$ coincides with the boundary of $A$ along the temporal evolution. The area functional is given by
\begin{equation}\label{areaaction}
\mathcal{A}={\rm Area}(\gamma_A)=\ell_\perp\int_{-\ell/2}^{\ell/2}dx\frac{\tilde{f}(z,v)}{z^2}\sqrt{\tilde{f}(z,v)^2-\tilde{g}(z,v)v'^2-2v'z'} \ ,
\end{equation}
where $'\equiv d/dx$. Since there is no explicit $x$-dependence, there is a corresponding conservation equation given by
\be\label{coneq}
\tilde{f}^2(z,v) - \tilde{g}(z,v) v'^2 - 2 v' z' = \tilde{f}^6(z,v)\left(\frac{z_*}{z}\right)^{4} \ .
\ee
In this expression, $z_*$ is defined through $z(0)=z_*$. Now, the right hand side of \eqref{efcoords}  in empty AdS is identically zero.  In our case it is of  order $\cO(\epsilon^4)$, which we can safely neglect since we are working to order $\cO(\epsilon^2)$. Thus, we have
\be\label{const}
v'+\frac{z'}{\tilde{g}(z,v)}\sim 0 \ .
\ee

Combining (\ref{coneq}) and (\ref{const}) we get
\be\label{zprime}
z'=\sqrt{\left(\tilde{f}^4(z,v)\left(\frac{z_*}{z}\right)^4-1\right)\tilde{f}(z,v)^2 \tilde{g}(z,v)} \ .
\ee
In practice we solve the above system as follows. First, we rewrite (\ref{const}) as
\be
\dot{v}+\frac{1}{\tilde{g}(z,v)}=0 \ ,
\ee
where $\dot{\;}\equiv d/dz$ and we solve for $v(z)$ subject to the boundary condition: $v(z_0)=t_{\rm boundary}$. With this function at hand and given a value of $z_*$, we can then obtain a solution for $z(x)$ by direct integration of (\ref{zprime}). However, note that this last step is not necessary if we only want to extract the values of $\ell$ and $\mathcal{A}$ for a given $z_*$. More specifically, these values can be obtained from
\be
\ell=2\int_{z_0}^{z_*}\frac{dz}{z'} \  \quad {\rm and} \quad \mathcal{A}=2\ell_\perp\int_{z_0}^{z_*}dz\frac{\tilde{f}(z,v)^4z_*^2}{z^4 z'} \ , \label{onshellA}
\ee
respectively. In particular, the last equation in (\ref{onshellA}) arises upon substituting (\ref{const})-(\ref{zprime}) in (\ref{areaaction}) and then changing the integration variable $dx\to dz/z'$.

On the other hand, the area in (\ref{onshellA}) is divergent as the UV cut-off $z_0\to 0$ and must be regularized. The divergence comes from the fact that the volume of any asymptotically AdS background is infinite and the spatial surface $\gamma_A$ reaches the boundary. In particular, near the boundary it is clear that $\tilde{f}(z\to0)\to1$, $z'(z\to0)=z_*^2/z^2$ and therefore
\be
\mathcal{A}_{\mathrm{div}} = 2\ell_\perp\int_{z\sim z_0}\!\! \frac{dz}{z^{2}} = \frac{2}{z_0} \ .
\ee
Subtracting this divergence, we obtain the finite term of the area which is the quantity we are interested in:
\be\label{areareg}
\mathcal{S}\equiv \mathcal{A}-\mathcal{A}_{\mathrm{div}}=2\ell_\perp\left(\int_{z_0}^{z_*}dz\frac{\tilde{f}(z,v)^4z_*^2}{z^4 z'}-\frac{1}{z_0}\right) \ .
\ee
%

\subsection{Toy model and choice of parameters}

Before proceeding to the numerical results, we must specify the system we want to study. First, notice that at $\mathcal{O}(\epsilon^4)$ the computation of entanglement entropy does not rely on the specific form of the coupling $h(\phi)$. Then, for the purposes of this section it is enough to set $h(0)=1$ and $\dot{h}(0)=0$, as required by the perturbative solution. For the scalar profile, we choose for simplicity a Gaussian function of the form
\be\label{gaussian}
\phi_1(v)=\lambda e^{-v^2/v_0^2} \ ,
\ee
where $\lambda$ and $v_0$ are numerical parameters that control the amplitude and the width. With this choice, the leading-order amplitude of the dilaton scales as $\phi\sim\epsilon\lambda$. Note that the perturbative expansion requires each order in \label{weakanstaz} to be at most of order $\mathcal{O}(1)$ in $\epsilon$. Here we have introduced the extra parameter $\lambda$ for convenience, which could yield $(\frac{\tilde{C}_2}{m})\sim\mathcal{O}(1)$ with a proper choice of its numerical value. We will come back to this point below. One advantage of the above profile is that it can be integrated analytically to obtain the explicit form of $C_2(v)$, $C_4(v)$ and $P(v)$. A brief computation leads to:
\be\label{m1profile}
C_2(v)=\frac{\lambda^2}{8 v_0^6}\left[4 v e^{-2 v^2/v_0^2} \left(4 v^2-3 v_0^2\right)+3 \sqrt{2 \pi } v_0^3 \left(2-\mathrm{erf}(\sqrt{2} v/v_0)\right)\right]\ ,
\ee
\bea
C_4(v)=&&\!\!\!\!\!\!\!\!\frac{\lambda ^4}{1536 v_0^6}\bigg[12v e^{-4 v^2/v_0^2} \left(72 v^2+25 v_0^2\right) +108\sqrt{2\pi }v_0^3e^{-2 v^2/v_0^2} \left(1+\mathrm{erf}(\sqrt{2} v/v_0)\right) \nonumber\\
&&\!\!\!\!\!\!\!\!-128\sqrt{3\pi } v_0^3e^{-v^2/v_0^2} \left(1+\mathrm{erf}(\sqrt{3} v/v_0)\right)+9\sqrt{\pi }v_0^3\bigg(1+\mathrm{erf}(2 v/v_0)\bigg)\bigg]\\
&&\!\!\!\!\!\!\!\!-\frac{3\lambda ^2}{16} e^{-2 v^2/v_0^2} k^2 m \ ,\nonumber
\eea
and
\be
\begin{split}
P(v)=&\;\frac{\lambda^3}{288v_0^6}\bigg[-12ve^{-3 v^2/v_0^2}\left(12v^2+v_0^2\right) +27 \sqrt{2\pi} v_0^3e^{-v^2/v_0^2}  \left(1+\mathrm{erf}(\sqrt{2} v/v_0)\right) \\
&\;-16\sqrt{3\pi}v_0^3\left(1+\mathrm{erf}(\sqrt{3} v/v_0)\right)\bigg]-\frac{\lambda}{4} e^{-v^2/v_0^2}k^2m \ .
\end{split}
\ee
Here ${\rm erf}(x)$ denotes the error function. Furthermore, we only need derivatives of $\phi_1(v)$ and $P(v)$ to compute the analytic forms of $\tilde{f}(z,v)$ and $\tilde{g}(z,v)$ according to (\ref{eq:finalfinal})-(\ref{eq:finalfinal4}).

\begin{figure}[!htm]
$$
\begin{array}{cc}
  \includegraphics[angle=0,width=0.48\textwidth]{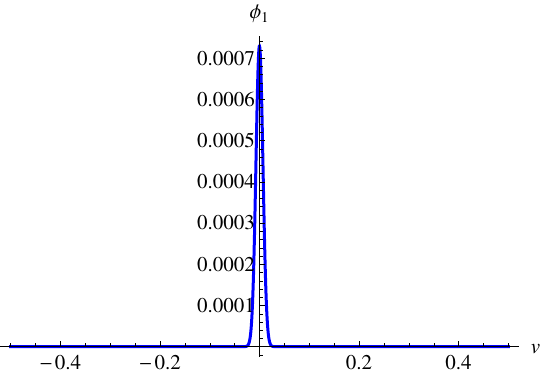} & \includegraphics[angle=0,width=0.48\textwidth]{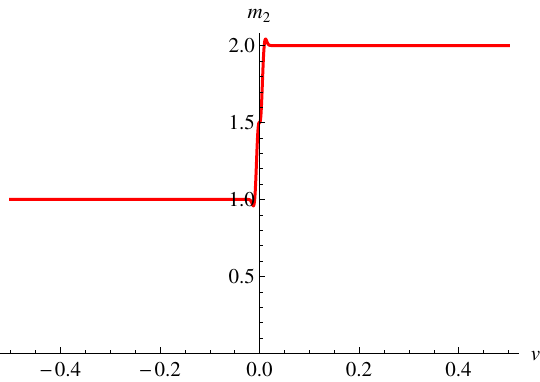}\\
  \!\!\!(a) & \!\!(b)
\end{array}
$$
\vspace{-6mm}
\caption{\small $(a)$ Gaussian profile for the scalar field $\phi_1(v)$ and $(b)$ mass function $m_2(v)$ according to (\ref{gaussian}) and (\ref{m1profile}), respectively. For the plots we chose $m=1$, $v_0=0.01$, $k=1$ and $\lambda$ given by (\ref{lamvalue}).}
\label{profiles}
\end{figure}

\begin{figure}[!htm]
\begin{center}
  \includegraphics[angle=0,width=0.65\textwidth]{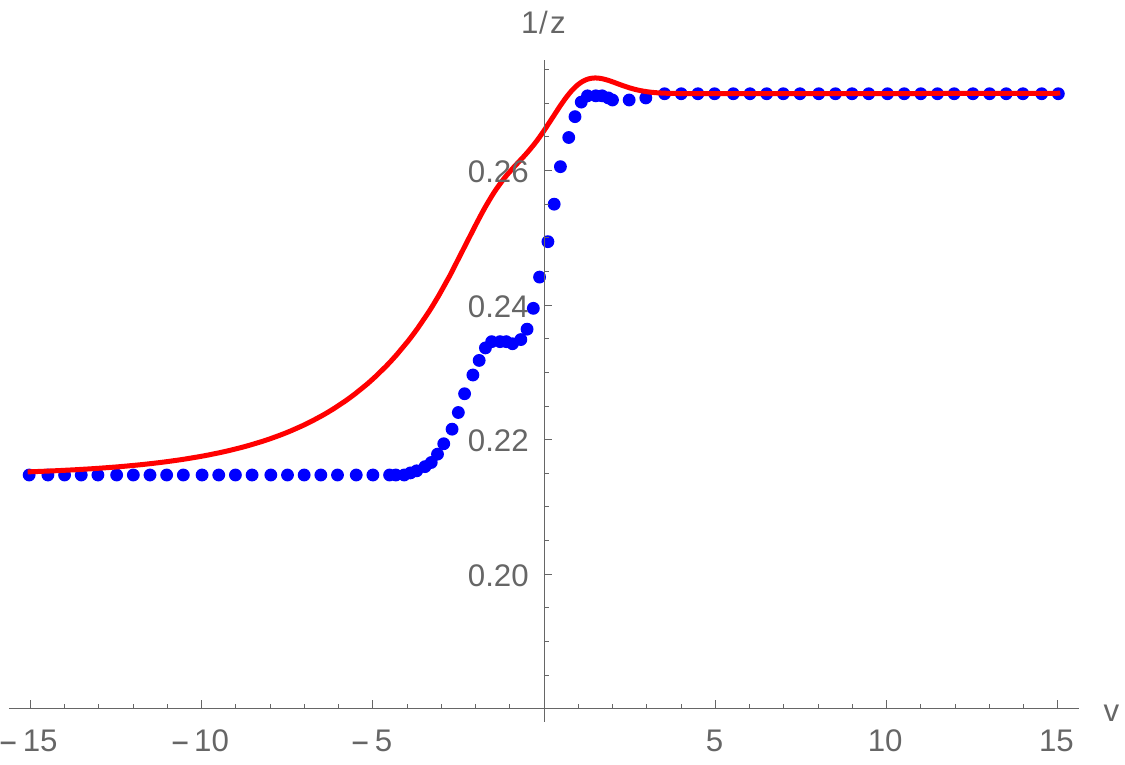}
\end{center}
\begin{picture}(0,0)
 \put(120,98){\small {\color{blue}$z_{\rm aH}(v)$}}
 \put(140,130){\small {\color{red}$z_{\rm H}(v)$}}
 \end{picture}
\vspace{-9mm}
\caption{\small Evolution of the the apparent horizon $z_{\rm aH}(v)$ (blue) vs. the event horizon $z_{\rm H}(v)$ (red) according to (\ref{eq:ahdef}) and (\ref{eq:ehdef}), respectively. For the plots we chose $m = 1$, $q=0.7$, $v_0 = 2$, $k=1$ and $\lambda$ given
by (\ref{lamvalue}). Notice that for this example we have chosen $v_0$ away from the thin shell limit in order to observe the non-monotonicity of the apparent horizon. If we let $v_0\to0$, this non-monotonic behavior is smoothed out and $z_{\rm aH}(v)$ approaches to a step function.\label{Horizons}}
\end{figure}
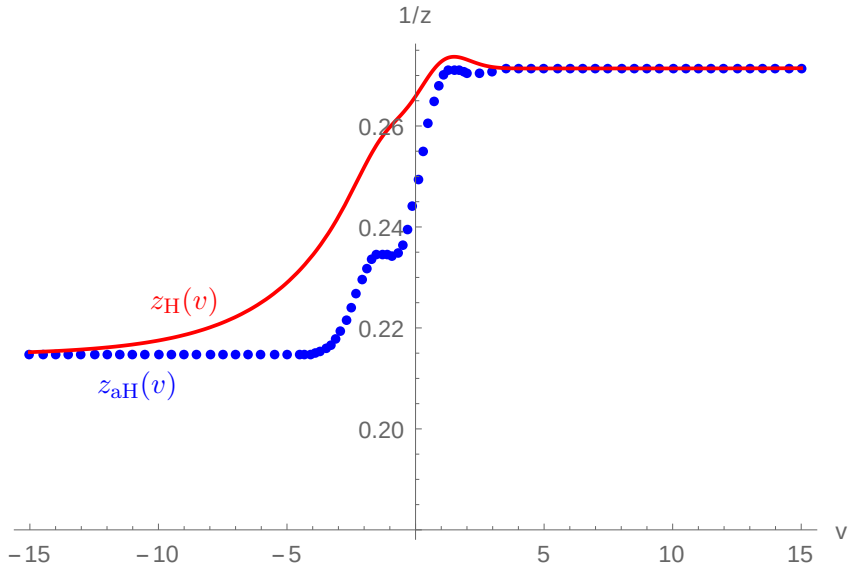

Next, we have to select values for the various parameters in order to be consistent with the perturbative expansion. First, we set $m=1$ and $v_0=0.01$. The first choice fixes a reference scale for the initial energy whereas the second guarantees that we are dealing with the thin shell limit. Then, we must fix $\lambda$ in order to have $m_2\sim\mathcal{O}(1)$ for the final state. Note that $\tilde{C}_2$ evaluates to
\be
\tilde{C}_2=\frac{3 \sqrt{2 \pi } \lambda ^2}{4v_0^3}\,.
\ee
Then, a good choice for $\lambda$ is
\be\label{lamvalue}
\lambda=\frac{2^{3/4} v_0^{3/2}}{\sqrt{3}\pi^{1/4}}\approx7.3\times10^{-4}\ ,
\ee
so that $\tilde{C}_2=1$. We further set $k=1$ so $\epsilon=\epsilon_1$ and $m_2=k^2m+\tilde{C}_2=2$. In Figure \ref{profiles} we plot both the gaussian profile for the scalar field and the mass function $m_2(v)\equiv k^2 m+\tilde{C}_2(v)$ for the parameters given above. The remaining functions $C_4(v)$ and $P(v)$ are of order $\mathcal{O} \left(10^{-7} \right)$ and  $\mathcal{O} \left(10^{-4} \right)$, respectively.
Note that the mass function does not increase monotonically in time, in stark contrast with the usual behavior of collapsing Vaidya geometries. Nevertheless, our field content is physically sensible and all the energy conditions are satisfied. Following the arguments of \cite{Caceres:2013dma}, then, we expect reasonable results for the thermalization process in the boundary theory.\footnote{We thank Esperanza Lopez for a discussion on this point.}  Another reason to argue that this must be true is that, although the \emph{apparent} horizon (\ref{eq:ahdef}) also shows this non-monotonic behavior, the \emph{event} horizon (\ref{eq:ehdef}), on the other hand, always increases along the temporal evolution.\footnote{If we truncate the metric at order $\mathcal{O}(\epsilon^4)$, we find that for $v_0\sim10$ or larger the event horizon also presents signs of non-monotonic behavior. This issue is corrected as we consider higher order corrections in $\epsilon$.} Figure \ref{Horizons} shows representative behaviors of these two quantities. At any rate, it is worth pointing out that such non-monotonic behavior disappears as we take the thin shell limit, which is the case we are focusing on. In this limit both $m_2(v)$ and $z_{\rm aH}(v)$ take the form of a step function.

The expansion parameter $\epsilon$ should be a small number. We observe that the first and second corrections to the functions $\tilde{f}(z,v)$ and $\tilde{g}(z,v)$ over AdS are of order $\mathcal{O}(\epsilon^2)$ and $\mathcal{O}(\epsilon^4)$, respectively. A reasonable choice for $\epsilon$ is then $\epsilon=0.1$. Finally, the value of $q$ can be tuned in order to vary the temperature and the chemical potential of the solutions. There are  two requirements for this quantity: $(i)$ we must impose that $q\sim\mathcal{O}(1)$ to be consistent with the perturbative expansion and $(ii)$ we cannot exceed the maximum value allowed in order to avoid a naked singularity at early times. Regarding this last point, recall that to our order of approximation, the initial state is characterized by
\be\label{temp1i}
T_i=\frac{3 \epsilon^{2/3}m^{1/3}}{4 \pi }\left(1-\frac{\epsilon ^{4/3}q^2 }{3m^{4/3}}\right) \ , \quad \mu_i=-\frac{\epsilon^{4/3}q}{m^{1/3}}\left(1+\frac{\epsilon ^{4/3}q^2}{6 m^{4/3}}\right) \ .
\ee
Therefore, from (\ref{temp1i}) it follows that the condition $(ii)$ for not having a naked singularity sets a maximum value
\be\label{cmax}
|q|_{\text{max}}=\frac{\sqrt{3}m^{2/3}}{\epsilon^{2/3}}\simeq 8.04 \ ,
\ee
above which $T_i<0$ and with no real roots of $g(z_{\rm H})=0$. Fortunately, notice that (\ref{cmax}) is also within the acceptable range for satisfying item $(i)$.

The final state, on the other hand, is characterized by
\be\label{temp1f}
T_f=\frac{3 \epsilon ^{2/3}m_2^{1/3}}{4 \pi}\left(1-\frac{\epsilon ^{4/3}q^2 }{3m_2^{4/3}}\right) \ , \quad \mu_f=-\frac{\epsilon ^{4/3} q}{m_2^{1/3}}\left(1+\frac{\epsilon^{4/3}q^2}{6 m_2^{4/3}}\right) \ ,
\ee
where
\be
m_2=m+\frac{3 \sqrt{2\pi } \lambda^2}{4 v_0^3}\ .
\ee
It is easy to check that for values of $q$ in the range allowed by (\ref{cmax}), the final states are also free of naked singularities. Of course, this is directly related to the fact that the mass of the black hole is increased while the charge is kept fixed.

Now, we have both temperature and chemical potential in our system and we need to construct a dimensionless ratio which would be our tunable parameter. Notice that both temperature and chemical potential has inverse length dimensions. Then, we can consider \cite{Caceres:2012em}
\begin{eqnarray} \label{ratio}
\chi \equiv \frac{1}{4\pi} \left(\frac{\mu_f-\mu_i}{T_f-T_i} \right)
\end{eqnarray}
to be the relevant parameter that we will vary. In practice, we can vary $\chi$ by tuning the parameter $q \in [-8.04,8.04]$. In particular, it is found that for this range of values, $\chi(q) \in [-0.42,0.42]$ and increases almost linearly with $q$. Thus, within the perturbative approximation and for the choices of the various parameters, we are restricted to small values of $\chi$ as compared to \cite{Caceres:2012em}.

\subsection{Regimes of thermalization}

\begin{figure}[!htm]
$$
\begin{array}{cc}
  \includegraphics[angle=0,width=0.48\textwidth]{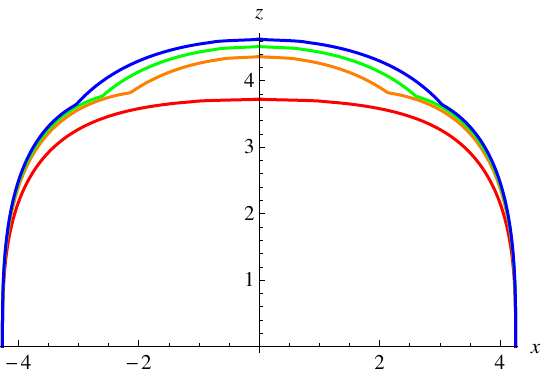} & \includegraphics[angle=0,width=0.48\textwidth]{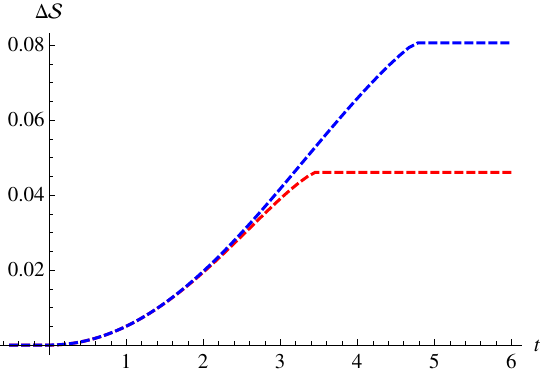}\\
    \!\!\!(a) & \!\!(b)
\end{array}
$$
\vspace{-6mm}
\caption{\small $(a)$ Sample embedding functions $z(x)$ for $\chi=0.2$ ($q\simeq3.651$), $4\pi \Delta T \ell=2$ ($\ell\simeq8.524$) and $t=\{5.6,6.8,8,9.2\}$ from top to bottom. $(b)$ Entanglement growth $\mathcal{S}(t)$ for $\chi=0.002$ (blue) and $\chi=2$ (red) with fixed $4\pi \Delta T \ell=1$. In both plots we have set the AdS radius to unity $L=1$.}
\label{EEgeo}
\end{figure}

Now we will discuss the numerical results. In Figure \ref{EEgeo} $(a)$ we plot sample solutions for the embedding functions $z(x)$ for a fixed $\ell$ as the boundary times $t$ is varied. Some of them cross the shell, located at $v=0$, and refract. This refraction is suppressed by a factor of $\epsilon^2$ given that the energy-momentum of the shell is itself of order $\mathcal{O}(\epsilon^2)$. Nevertheless, the aforementioned effect is noticeable to the naked eye for large enough distances, $\left(4\pi \Delta T\right) \ell\gtrsim 1$. We also show in $(b)$ the behavior of entanglement entropy as a function of time, for a fixed distance $\ell$. In order to compare the results for various values of $\chi$ we subtract the entropy of the initial state $\Delta S(t)=S(t)-S(-\infty)$, and focus on the entanglement growth over time.\footnote{The divergent piece of the entanglement entropy is independent of the legth. Here we are subtracting also a finite piece that depends on $\ell$ but does not display temporal evolution. With this subtraction, the entanglement entropy $\Delta S(t)$ starts at zero in the infinite past for all values of $\ell$.} We note some general properties of the behavior of $\Delta S(t)$ as we change the value of $\chi$. Qualitatively, our results agree with those of \cite{Liu:2013iza,Liu:2013qca} (see also \cite{Alishahiha:2014cwa, Fonda:2014ula, Alishahiha:2014jxa}) obtained in the context of Vaidya geometries. At early times, {\it i.e.}~the ``pre-local-equilibration'' regime, the evolution is almost quadratic in time and is weakly dependent on the size $\ell$. This stage is followed by a linear growth phase at intermediate times and, finally, a saturation is reached at late times, $t\geq t_{\mathrm{sat}}$, where the entropy abruptly flattens out.

\begin{figure}[!htm]
\begin{center}
  \includegraphics[angle=0,width=0.65\textwidth]{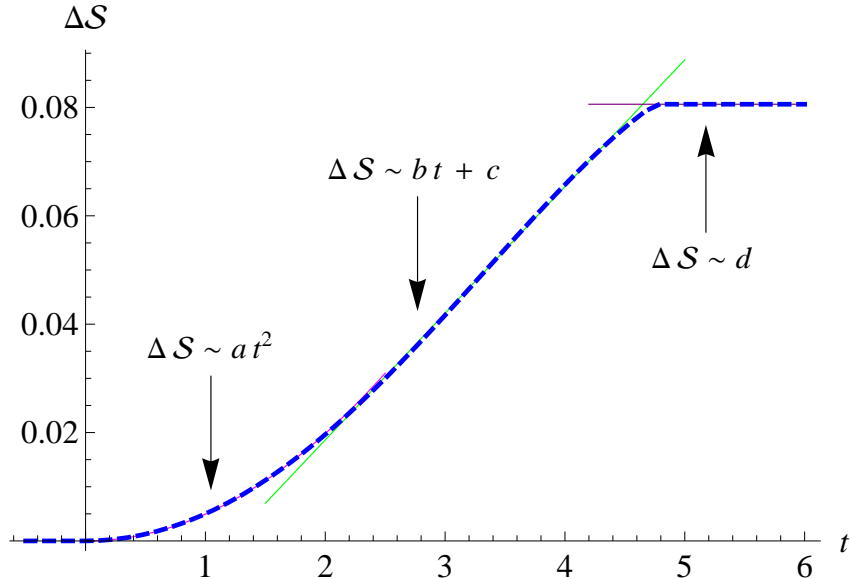}
\end{center}
\vspace{-5mm}
\caption{\small Functional dependence of entanglement entropy as a function of time in the various regimes of thermalization. For this example we fixed $\chi=0.002$ and $4\pi \Delta T \ell=1$, and the best fits yielded the following constants: $a=0.00496$, $b=0.02338$, $c=-0.02813$ and $d=0.08059$.  The time here is measured in units of the AdS radius which has been set to unity $L = 1$.\label{EEregimes}}
\end{figure}

Let us explain these regimes in more details. First, in Figure \ref{EEregimes}, we have shown the functional dependence of entanglement entropy growth in various regimes. Following \cite{Liu:2013iza} we can introduce a ``local equilibrium scale",  denoted by $\ell_{\rm eq}$, which means that a local thermodynamic description applies at length scales$\sim \cO(\ell_{\rm eq})$ even though the global description is out-of-equilibirum. For early times, which can be represented by the regime $t \ll \ell_{\rm eq}$, the rate of entanglement entropy growth is expected to be proportional to the area of the entangling surface. Furthermore, we can assume that the growth is proportional to a characteristic energy scale of the system. Note that, at the initial stage, we have the energy of the initial state, denoted by $E_{\rm initial}$, and the energy which is being injected to induce the dynamics, denoted by $\Delta E$. Thus the typical energy-scale should be identified as
\begin{eqnarray}
E_{\rm typical} = {\rm max} \left \{ E_{\rm initial}, \Delta E \right \} \ .
\end{eqnarray}
{\it A priori}, $E_{\rm initial}$ and $\Delta E$ are independent. However, in our case, we already demanded $E_{\rm initial} \sim \Delta E$ and thus, using dimensional analysis, we can arrive at
\begin{eqnarray}
\Delta S(t) = \left(\alpha_1 E_{\rm initial} \cA\right) t^2 = \left( \alpha_2 \Delta E \cA \right) t^2 \ , \label{quadratic}
\end{eqnarray}
where $\alpha_{1,2}$ are two constants and $\cA$ denotes the area of the entangling surface.

On the other hand, in the regime $ t \gg \ell_{\rm eq}$, we have a notion of a thermal entropy density which is denoted by $S_{\rm thermal}$. If we further assume that the entanglement entropy is proportional to the area of the entangling surface and the local thermal entropy density, then by dimensional analysis we get
\begin{eqnarray}
\Delta S (t ) = \left(v_{\rm E} \cA S_{\rm thermal} \right) t \ , \label{linear}
\end{eqnarray}
where $v_{\rm E}$ is the entanglement production rate which has been analyzed in \cite{Liu:2013iza, Liu:2013qca}. Evidently, our numerical data agrees very well with the intuition outlined in (\ref{quadratic}) and (\ref{linear}). It is intriguing to further note that, although we are not starting from a vacuum state, the analysis of \cite{Liu:2013iza, Liu:2013qca} continues to hold.\footnote{The issue of state dependence was further studied in \cite{Fischler:2013fba} for the case of hyperbolic AdS-Vaidya black holes, finding qualitative agreement with the results of \cite{Liu:2013iza, Liu:2013qca}.} Finally, since we are considering the ``rectangular" shaped entangling surface, we expect that the saturation will be accompanied by an abrupt jump in the corresponding
extremal area geodesic.

Now we will discuss the scaling of thermalization time. From the time-evolution of entanglement entropy, we can extract $t_{\rm sat}$ for a given length $\ell$. Alternatively, we can define another time-scale $t_{\rm crit}$ as a measure of the thermalization time. Recall that the shell is densely peaked around $v=0$ for $v_0 \ll 1$. Thus, we can define $t_{\rm crit}$ to be the time at which the corresponding extremal surface grazes the shell at $v=0$. By definition from (\ref{efcoords}), we get
\begin{eqnarray}\label{criticalt}
t_{\rm crit} = \int_{z_0}^{z_*} \frac{dz} {\tilde{g} (z, v=0)} \ .
\end{eqnarray}
In practice, it is easier to extract $t_{\rm crit}$ rather than $t_{\rm sat}$.\footnote{Note, however, the approach to equilibrium is expected to be abrupt in the case of a rectangular entangling region\cite{Liu:2013iza}. Thus, strictly speaking, $t_{\rm crit}$ may not be the correct measure of thermalization time. However, in our case, we have checked that $t_{\rm sat}$ and $t_{\rm crit}$ exhibit qualitatively similar behaviour. Since we are only concerned with qualitative features, we do not attempt to make this more precise here.} Furthermore, in the limit $v_0\to0$ these two quantities are found to agree. For the case of finite thickness, they only differ by a factor of order $\mathcal{O}(v_0)$ so we expect similar results for the thermalization time as long as $v_0$ is not too large. We thus focus on the critical time $t_{\rm crit}$.
\begin{figure}[!htm]
$$
\begin{array}{cc}
  \includegraphics[angle=0,width=0.48\textwidth]{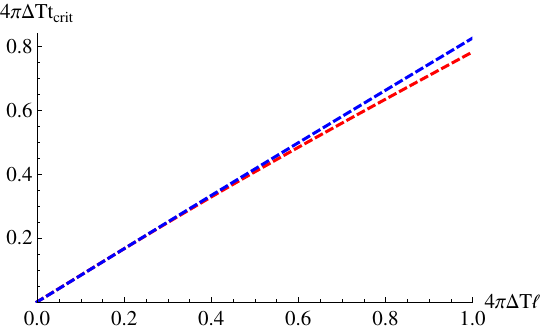} & \includegraphics[angle=0,width=0.48\textwidth]{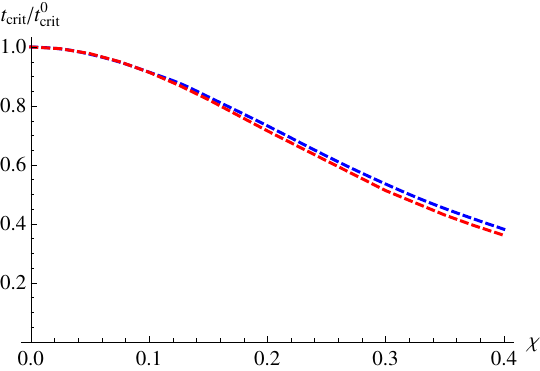}\\
    \!\!\!(a) & \!\!(b)
\end{array}
$$
\vspace{-6mm}
\caption{\small $(a)$ Critical time as a function of length for fixed $\chi=0.04$ (red) and $\chi=0.4$ (blue). $(b)$ Critical time as a function of $\chi$ for a fixed length $4\pi \Delta T\ell=0.1$ (red) and $4\pi \Delta T\ell=1$ (blue). The AdS radius has been set to unity $L=1$.}
\label{EEGrowth}
\end{figure}

Let us, however, clarify a further caveat regarding $t_{\rm crit}$. Precisely speaking, $t_{\rm crit}$ measure the time a null ray takes to reach the AdS-boundary starting from the bulk point $z_\ast$. Thus, for a given boundary length $\ell$, the corresponding extremal area geodesic will indeed graze the shell at $t_{\rm crit}$ provided the shell propagates at the speed of light. However, as can be checked from (\ref{EMmax}) and (\ref{EMscalar}), the matter field is non-null and thus propagates slower than the speed of light. Thus, in reality $t_{\rm crit}$ will serve as a lower bound for the actual thermalization time.

The dependence of $t_{\rm crit}$ with $\ell$ is shown in Figure \ref{EEGrowth} $(a)$, which approximates a linear growth similar to the one observed in \cite{Balasubramanian:2010ce, Balasubramanian:2011ur}. Generally, we can represent the dependence as
\begin{eqnarray}
\Delta T t_{\rm crit} =  A \left( \chi \right) \left( \Delta T \ell \right) + B \left( \chi \right) \left( \Delta T \ell \right)^{\alpha(\chi)}  \ ,
\end{eqnarray}
where $A(\chi)$ represents the slope of the linear regime, {\it i.e.}~the velocity at which thermalization propagates in the system. Numerically we find
\begin{eqnarray}
A(\chi) & = & 0.83 \quad {\rm for \, \, blue} \\
& = &   0.79 \quad {\rm for \, \, red} \ ,
\end{eqnarray}
which implies that $t_{\rm crit}$ sets a super-critical (correspondingly a faster than speed of light propagation) time-scale.\footnote{Recall that, $t_{\rm crit}$ sets a lower bound on the actual thermalization time. Therefore, a super-critical $t_{\rm crit}$ does not correspond to a superluminal thermalization. This essentially means that the actual thermalization time is always larger than $t_{\rm crit}$, which should satisfy causality constraints\cite{Balasubramanian:2011ur}. To extract the actual thermalization time, one needs to carry out a thorough analyses of the corresponding extremal-area surfaces, which we leave for future work. Here we merely want to demonstrate that the lower bound set by $t_{\rm crit}$, which nevertheless is below the causality constraint, decreases with increasing chemical potential.} At larger length-scales, the deviation from linearity is characterized by the second term above, where $\alpha(\chi)$ is an index which can, in principle, depend on $\chi$. 

On the other hand $t_{\rm crit}$ monotonically decreases with $\chi$ for fixed $\ell$ (in the allowed range for $\chi$), which means that thermalization is faster in the presence of a chemical potential. We have shown a representative behaviors in Figure \ref{EEGrowth} $(b)$. Our findings agree qualitatively with the results of \cite{Caceres:2012em}, in the regime of validity of our solutions.

Let us now comment on what may happen beyond the validity of our perturbative solutions. It is unlikely that the thermalization time keep decreasing with increasing chemical potential. Hence there are two possibilities: (i) thermalization time plateaus or (ii) thermalization time eventually turns around and starts increasing with increasing chemical potential. Either way, it implies a qualitatively different scaling behaviour of thermalization time in two regimes: when $\chi \ll1 $ and when $\chi \gg 1$, {\it i.e.}~which has an obvious interpretation as a ``classical" and a ``quantum" regime respectively. The latter observation is similar to the one made in \cite{Caceres:2012em}.

\subsection{A remark on the scrambling time}

Before closing this section, we wish to make a few comments.\footnote{We thank Diego Trancanelli for bringing this point to our attention.} Our initial state is thermal, and thus is represented in the gravity description by an AdS black hole. It is generally believed that black holes are endowed with the special property of ``fast scrambling'' \cite{Sekino:2008he,Susskind:2011ap}. In particular, this implies that the time scale associated to the process of thermalization grows logarithmically with the number of degrees of freedom of the system
\be\label{scrtime}
t_*\sim\beta \log N_c \ ,
\ee
where $\beta$ is independent of $N_c$. In our case, we can count $N_c$ by evaluating the thermal entropy of the initial state $S_{\rm Thermal}^{i}$, or of the final state $S_{\rm Thermal}^{f}$, or of their difference $\Delta S_{\rm Thermal}$.
\begin{figure}[!htm]
\begin{center}
  \includegraphics[angle=0,width=0.65\textwidth]{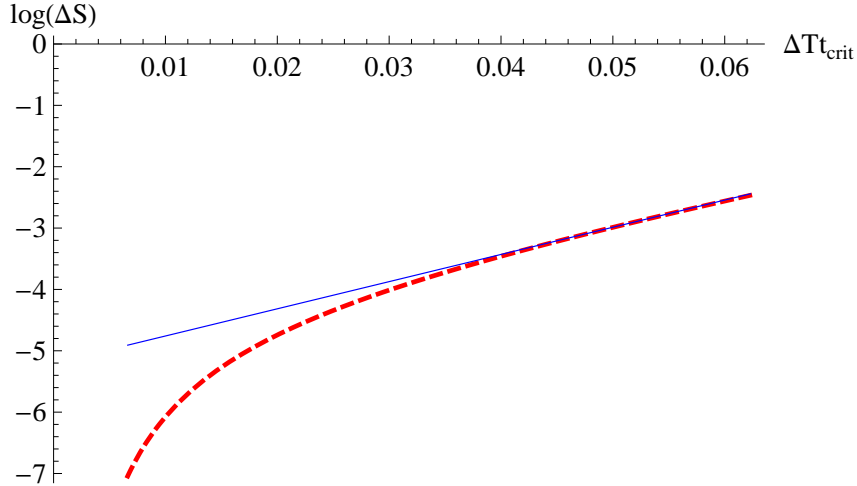}
\caption{\small Behavior of entanglement entropy vs. saturation time for a fixed $\chi=0.04$. The plot is generated by joining the points $\{\Delta S (\ell),\Delta T t_{\rm crit}(\ell)\}$ corresponding to different values of $\ell$. The blue line represent the best fit in the farthermost region, for which the values of length are of order $\Delta T\ell\sim1$. \label{scrambling}}
\end{center}
\end{figure}

Now, in order to make contact with the scrambling property of the black hole, we have to focus on the regime $\Delta T\ell\gg1$ where entanglement entropy approaches the thermal entropy, $S(\Delta T\ell\gg1)\sim S_{\rm Thermal}$. In this same limit, we can identify $t_{\rm crit}\sim t_*$ since $t_{\rm crit}(\Delta T\ell\gg1)$ serves as a measure of ``global'' equilibration. In Figure \ref{scrambling} we show the behaviour of $\log \Delta S$ as a function of the $\Delta T t_{\rm crit}$ for a fixed value of chemical potential $\chi$. To generate the plot, we vary the length $\ell$ to generate pair of points $\{\Delta S (\ell),\Delta T t_{\rm crit}(\ell)\}$ and then we join them. Larger values of $\ell$ correspond to larger $t_{\rm crit}(\ell)$ for a fixed $\Delta T$ (see Figure \ref{EEGrowth}), so we are interested in the rightmost part of Figure \ref{scrambling}. It is worth pointing out that, due to limitations of the numerical accuracy and the validity of our perturbative solution, we can only go as far as $\Delta T\ell\sim1$. This is because in order to increase the length of the boundary strip we have to reduce $z_{\rm H}-z_*$ accordingly, which needs to be fine-tuned to high precision in order to increase $\Delta T\ell$. However, it is intriguing that in the region around $\Delta T\ell\sim1$ the curve smoothly approaches a straight line, in concordance with (\ref{scrtime}). It would be remarkable if this statement holds true as $\Delta T\ell$ is increased. If it does, this may lead to important clues on how the effective degrees of freedom interact towards the process of thermalization\cite{Lashkari:2011yi,Edalati:2012jj}. It will be interesting to investigate this issue further and make contact with other approaches within the framework of AdS/CFT (see, \emph{e.g.} \cite{Shenker:2013pqa,Leichenauer:2014nxa}).

\subsection{Evolution of $L(v)$: a local observable}

In section \ref{sec:level2}, we remarked that we can observe a non-trivial evolution of local observables in our case. A natural local observable is the vacuum expectation value (VEV) of the marginal operator which we are quenching here. To extract the precise VEV, we need to carefully perform holographic renormalization of our perturbative solution in (\ref{evo1})-(\ref{evo4}). However, here we merely want to illustrate this time-evolution, and the entire machinery of holographic renormalization is redundant for our purposes.

\begin{figure}[!htm]
\begin{center}
  \includegraphics[angle=0,width=0.65\textwidth]{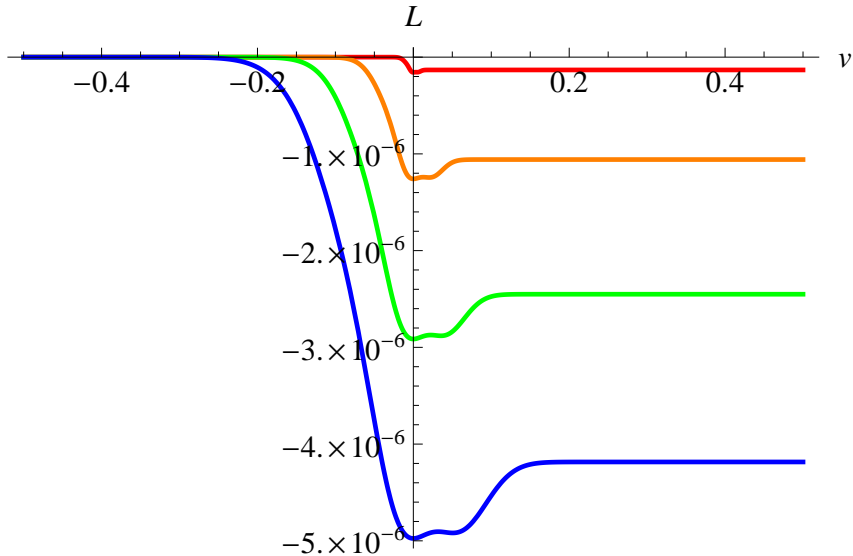}
\end{center}
\vspace{-5mm}
\caption{\small Time evolution of the normalizable mode of the scalar field $L(v)$ according to (\ref{Lveq}). Both vertical and horizontal axes are measured in units of the AdS radius, which has been set to unity. For the plots we have chosen parameters $m=1$, $k=1$ and $\epsilon=1/10$. The different colors correspond to different widths for the non-normalizable mode of the scalar profile: $v_0=0.01$ (red), $0.04$ (orange), $0.07$ (green) and $0.1$ (blue), respectively.\label{locobs}}
\end{figure}
Thus, in figure~\ref{locobs} we show the time-evolution of $L(v)$, which is given in equation (\ref{Lveq}).\footnote{Note that, the precise VEV of the marginal operator, if we were to perform holographic renormalization, would differ from $L(v)$ by derivative terms of $\phi_1(v)$, which vanish at the initial and the final states.} Note that, the relatively small magnitude of the vertical axis is a consequence of the choice of the parameters that we have previously made. It is clear that the evolution occurs at a time-scale of $\cO(\delta t) = \cO(v_0)$.

 \setcounter{equation}{0}
\section{\label{sec:level7} Conclusions}

We have considered here the thermal quench of a marginal operator in a prototypical large $N_c$-gauge theory. In order to establish the scaling of the thermalization time on a more robust ground, there are many directions which we intend to explore in near future.

First, we can generalize our perturbative analysis when the quench is being carried out for a relevant operator. This will correspond to introducing a non-trivial scalar potential and it will be interesting to check to what extent our observations are universal. In turn, it is easier to embed such effective gravity descriptions within $11$-dimensional supergravity. Subsequently it will also be interesting to understand the specific embedding of our effective gravity description, in which the scalar potential is vanishing, in string theory. Note that, our model is unlikely to be realized within ABJM, since the latter does not have any marginal scalar-deformation.\footnote{All scalars have a non-vanishing mass around the corresponding AdS-fixed point.}

To test the robustness of our qualitative observation, it will be interesting to model a similar dynamical background in the presence of more than one global U$(1)$-charge, by {\it e.g.}~coupling the STU-model with a neutral massless scalar.\footnote{Note that charged black hole solutions for STU-models possess a very rich phase diagram\cite{Chamblin:1999tk}.} The quench of this scalar field will correspond to the quench of a marginal or relevant operator in the dual field theory. It will be very interesting to check how thermalization behaves in this case.

We have considered only $(3+1)$-dimensional bulk theory which corresponds to a $(2+1)$-dimensional boundary theory. It is well-known that the dynamics in asymptotically locally AdS$_{d+1}$ differs qualitatively for even vs.~odd $d$, see {\it e.g.}~\cite{Bhattacharyya:2009uu}. Thus generalizing our results for the asymptotically AdS$_5$-background will be an interesting avenue to pursue.

Intuitively a generic charged-system is much richer than the neutral case and one can consider various possibilities. In this article, we considered the quench by a neutral scalar. One can also consider a charged-scalar field. In the latter, depending on the temperature of the system, the ground state of the system can be described by a Reissner-Nordstorm black hole or a hairy black hole, corresponding to a ``normal" phase and a superconducting phase of the field theory. Clearly, this corresponds to a more complicated dynamical process and it will be worthwhile to check how much physics can be accessed {\it via} a perturbative approach analogous to what we have adopted here.

Generalizing on this theme, it will be interesting to catalogue the possibilities in which Einstein equations admit such perturbative analyses. A bottom-up approach can be followed with various matter content in an Einstein-gravity theory with a negative cosmological constant. We can further relax the asymptotically AdS-condition, by using an asymptotically Lifshitz or Hyperscaling-Violating geometries.

The importance of a complete numerical evolution can hardly be overestimated. It is very crucial that eventually we have access to the entire evolution process in the full parameter space. As outlined above, there are numerous physically inequivalent real-time phenomena that can be captured within Einstein gravity with Maxwell and (charged or uncharged) scalar fields in various dimensions, which needs extensive numerical explorations. This is a long-term goal which we leave for future.

\section*{Acknowledgements}

We are grateful to Alex Buchel, Sumit Das, Willy Fischler, Stephen Green, Vadim Kaplunovsky, Luis Lehner, Esperanza Lopez, Leopoldo Pando Zayas and Diego Trancanelli for enlightening discussions at different stages of this project. We specially thank Alex Buchel, Sumit Das, Esperanza Lopez, Leopoldo Pando Zayas and Diego Trancanelli for very useful comments on the manuscript. This work is supported by Mexico's National Council of Science and Technology (CONACyT) grant CB-2008-01-104649 (EC), in part by a Simons Fellowship and in part by ERC Starting Grant HoloLHC-306605 (AK), in part by the NSF under Grant PHY-1316033 and by the Texas Cosmology Center (EC and JFP), Government of Canada through Industry Canada and by the Province of Ontario through the Ministry of Research and Innovation (JFP), and in part by European Union's Seventh Framework Programme under grant agreements (FP7-REGPOT-2012-2013-1) no 316165, the EU-Greece program ``Thales" MIS 375734 and CYCU under the grant MOST 103-2811-M-033-002 (DLY). DLY was also co-financed by the European Union (ESF) and Greek national funds through the Operational Program ``Education and Lifelong Learning" of the National Strategic Reference Framework (NSRF) under ``Funding of proposals that have received a positive evaluation in the 3rd and 4th Call of ERC Grant Schemes". AK further acknowledges the warm hospitality at the Aspen Center for Physics, where he was supported in part by NSF Grant No.~PHYS-1066293 and in part by the Simons Foundation; and the Mainz Institute for Theoretical Physics for its hospitality and support, where significant parts of this work were done.

\appendix
\numberwithin{equation}{section}

\section{Series solution up to 8th order in amplitude}

In section \ref{sec:epsilonexpansion} we presented the perturbative solution up to order $\epsilon^4 $. In this appendix we will present  the explicit form of the solution at higher  orders of $\epsilon$, where specially the non-trivial contributions coming from the coupling of the neutral scalar with the gauge field become transparent.

Recall that  in \eqref{eq:epsilonexpansion} we gave
 the formal structure of the functions entering the $\epsilon$ expansion (\ref{weakanstaz1})\,-\,(\ref{weakanstaz4}),
 \begin{align}\label{eq:epsilonexpansion-app}
		& \phi_{2n+1}(z,v) =\sum_{k=0}^{2n-2} z^{2n+1-k} \phi^k_{2n+1}(v) \ , \qquad n\ge2 \nonumber\\
		& f_{2n}(z,v) = \frac{1}{z} \sum_{k=0}^{2n-4}z^{2n-k} f^k_{2n}(v) \ , \qquad n\ge3 \\
		& g_{2n}(z,v) = z  C_{2n}(v)  + \frac{1}{z} \sum_{k=0}^{2n -3} z^{2n-k} g_{2n}^k(v) \ , \qquad  n\ge3 \nonumber\\
		& A_{v_{2n}}(z,v) = \mu_{2n}(v) + \sum_{k=0}^{2n-2} z^{2n-1-k} A_{v_{2n}}^k(v) \ .
		\qquad n\ge 3\nonumber
	\end{align}

The solution up to order $\epsilon^4$ involves the functions  $C_2(v), P(v), C_4(v) $ and $a_4(v)$  whose explicit form was given  in (\ref{eq:C2})\,-\,(\ref{eq:a4}). Since these functions  also enter  the higher order expansion  we quote them here again for sake of completness.
\begin{align}
		& C_2(v) = -\frac{1}{2} \int _{-\infty}^ v \phi_1'(x) \phi_1'''(x) dx  \ , \\
		& C_4(v) = \frac{3}{8} \int_{-\infty}^v  \phi_1'(x) \left( - \phi_1'(x)^3 + \int_{-\infty}^x( B(y)-m k^2 \phi_1'(y))dy \right) dx \ , \\
		& B(x) = \phi_1'(x) \left( C_2(x) + \phi_1'(x) \phi_1''(x) \right) \ , \\
		& P(v) = \frac{1}{4}\int_{-\infty}^v \left(- m k^2  \phi_1'(x)+ B(x) \right) dx  \  , \\
		& a_4(v) = - q k^2  \int _{-\infty}^ v \phi_1 (x) \phi_1'(x) \ddot{h}(0) dx  \ .
	\end{align}
where $k=\epsilon_1/\epsilon\sim \mathcal{O}(1).$  We  now proceed to write the solution to order $\epsilon^5,\, \epsilon^6, \, \epsilon^7$ and $\epsilon^8$.

\subsubsection*{Order $\epsilon^5$:}

$f_5(z,v)=g_5(z,v)=A_{v_{5}}(z,v)=0$.

\begin{align}
\bullet\, \phi_{5}(z,v)&= z^5 \phi_5^0 (v) + z^4 \phi_5^1(v) + z^3 \phi_5^2(v) \ ,
\end{align}
where,
\begin{align}
\phi_5^0(v)&=\frac{1}{192}\int_{-\infty}^v(-216 (k^2 m  +C_2(x))P(x) + 3 \fp(x)^3(k^2 m + C_2(x)) +\nonumber \\& \qquad \qquad\qquad (84 P(x)\fp(x) + 5 \fp(x)^4) \fpp(x))dx \ , \\
\phi_5^1(v)&=\frac{1}{144}\int_{-\infty}^v (240\phi_5^0(v) + 24 k^4 q^2 \fp(x) -120 P(x) \fp(x)^2 + 7 \fp(x)^5 + 24 k^4 q^2 \fp(x) h''(0))dx \ , \nonumber\\ \\
\phi_5^2(v)&=\frac{1}{4}\int_{=\infty}^v \left(4 \phi_5^1(x) + C_4(v) \fp(v)+k^4 q^2 \f(x) h''(0) \right)dx \ .
\end{align}

\subsubsection*{ Order $\epsilon^6$:}

Here we have $\phi_{6}(z,v)=0$. The remaining functions take the form:

\begin{align}
\bullet\, g_6(v) &= z C_6(v) +\frac{1}{z} \left( g_6^0(v) z^6 + g_6^1(v) z^5 + g_6^2(v) z^4 + g_6^3(v) z^3 \right) \ ,
\end{align}

with
\begin{equation}
C_{6}(v)= \int_{-\infty}^{v} \frac{3}{2} \phi_5^{2}(x)\fp(x) dx \ ,
\end{equation}

\begin{multline}
g_6^0(v)= -\frac{1}{4} \left( \frac{1}{20}k^2 m P(v) \fp(v) +\frac{1}{20} C_2(v) P(v) \fp(v) + \frac{13}{480} k^2 m \fp(v)^4 +\frac{13}{480} C_2(v) \fp(v)^4 \right. \\
\left. -\frac{2}{3} \phi_5^0(v)\fpp(v) -\frac{1}{24}P(v)\fp(v)^2\fpp(v) +\frac{17}{1440}\fp(v)^5\fpp(v) \right) \ ,
\end{multline}
\begin{multline}
g_6^1(v)=-\frac{1}{3} \left( \frac{9}{20} P(v)^2 -\frac{1}{2}\phi_5^0(v)\fp(v) -\frac{9}{40} k^4 \, q^2\, \fp(v)^2 -\frac{1}{5}P(v)\fp(v)^3 +\frac{19}{960}\fp(v)^6 + \right. \\ \left. \frac{3}{20} k^4 \, q^2\, \fp(v)^2 h''(0)
-\frac{3}{5} \phi_5^1(v)\fpp(v) \right) \ ,
\end{multline}
\begin{multline}
g_6^2(v) = -\frac{1}{2} \left( -\frac{1}{2} \phi_5^1(v) \fp(v) -\frac{1}{8} C_4(v)\fp(v)^2 +\frac{3}{8} k^4\, q^2\, \f(v) \fp(v) h''(0) -\frac{1}{2}\phi_5^2(v) \fpp(v) \right) \ ,
\end{multline}
\begin{equation}
g_6^3(v)=k^2\, q^2\, a_4(v) +\frac{1}{2} \phi_5^2(v)\fp(v) +\frac{1}{4} k^4\, q^2\, \f(v)^2 h''(0) \ .
\end{equation}

\begin{align}
\bullet \,  f_6(v)&= \frac{1}{z} \left( f_6^0(v) z^6 + f_6^1(v) z^5 + f_6^2(v) z^4 \right) \ ,
\end{align}
where,
\begin{equation}
f_6^0(v)=\frac{1}{60} \left(-\frac{9}{2}P(v)^2 - 5\phi_5^0(v)\fp(v)+\frac{7}{16}P(v)\fp(v)^3-\frac{1}{768}\fp(v)^6\right) \ ,
\end{equation}
\begin{equation}
f_6^1(v)=-\frac{1}{10}\phi_5^1(v)\fp(v) \ ,
\end{equation}
\begin{equation}
f_6^2(v)=-\frac{1}{8}\phi_5^2(v)\fp(v) \ .
\end{equation}

\begin{align}
\bullet \,  A_{v_6}(v) & = \mu_6(v) + A_{v_6}^0(v) z^5 + A_{v_6}^1(v)z^4 + A_{v_6}^2(v) z^3 + A_{v_6}^3(v) z^2 + A_{v_6}^4(v)z \ ,
\end{align}
with
\begin{multline}
A_{v_6}^0(v) = -\frac{1}{240}\big(-12k^2\, q\, P(v) \fp(v) - 2\, k^2\, q\, \fp(v)^4 + 48 \, k^2\, P(v) \fp(v) h''(0) + 6 \, k^2\, q\, \fp(v)^4 h''(0)- \\
12\, k^2\, q\, \fp(v)^4 h''(0)^2 + 2\, k^2\, q\, \fp(v)^4 h^{(4)}(0) \big) \ ,
\end{multline}
\begin{multline}
A_{v_6}^1(v) = -\frac{1}{144} \left(36 \, k^2\, q\, \f(v) P(v) h''(0) + 9\, k^2\, q\, \f(v)\fp(v)^3 h''(0)- 36 k^2\, q\, \f(v)\fp(v)^3 h''(0)^2 + \right. \\
\left. 6 \, k^2\, q\, \f(v)\fp(v)^3 h^{(4)}(0) \right ) \ ,
\end{multline}

\begin{multline}
A_{v_6}^2(v) = -\frac{1}{72}\big(-6 a_4(v)\fp(v)^2 + 12 a_4(v) \fp(v)^2 h''(0) - 30\, k^2\, q\, \f(v)^2 \fp(v)^2 h''(0)^2 + \\
6\, k^2 \, q\, \f(v)^2 \fp(v)^2 h^{(4)}(0)\big) \ ,
\end{multline}

\begin{multline}
A_{v_6}^3(v) = -\frac{1}{24}\big(12 a_4(v)\f(v)\fp(v)h''(0)-6 k^2\,q\,\f(v)^3\fp(v)h''(0)^2+ 2\,k^2\,q\,\f(v)^3\fp(v)h^{(4)}(0)\big) \ ,
\end{multline}
\begin{equation}
A_{v_6}^4(v) = \frac{1}{6}\int_{-\infty}^v\f(x)\fp(x)(-6 a_4(x) h''(0) + 3k^2 \,q\, \f(x)^2 h''(0)^2 -k^2 \, q\, \f(x)^2 h^{(4)}(0)) dx \ .
\end{equation}

\subsubsection*{ Order $\epsilon^7$:}

$f_7(z,v)=g_7(z,v)=A_{v_7}(z,v)=0$.
\begin{equation}
\bullet\,  \phi_{7}(z,v)= z^7 \phi_7^0(v) + z^6 \phi_7^1(v) + z^5 \phi_7^2 (v) + z^4 \phi_7^3(v) + z^3 \phi_7^4(v) \ ,
\end{equation}
where

\begin{multline}
\phi_7^0(v)=\frac{1}{23040}\int_{-\infty}^v \left(-4800(k^2m +C_2(x))\phi_5^0(x)+456(k^2 m +C_2(x))P(x)\fp(x)^2+ 67(k^2 m+ C_2(x))\fp^5
+ \right. \\
\left. 10368 P(x)^2 \fpp(x) +8640 \phi_5^0(x) \fp(x)\fpp(x) + 1332 P(x) \fp(x)^3 \fpp(x) + 69 \fp(x)^6 \fpp(x)\right) dx \ ,
\end{multline}
\begin{multline}
\phi_7^1(v)=\frac{1}{7200}\int_{-\infty}^v\bigg(20160 \phi_7^0(x) + 1728 P(x)^2 \fp(x) + 12 P(x)(41\fp(x)^4 +60k^4q^2(6+h''(0)))-\\
 2880 \phi_5^1(x) ( 4 k^2 +4 C_2(x) -\fp(x)\fpp(x)) + \\
 \fp(x)^2 (-11280 \phi_5^0(x)+ \fp(x)(43\fp(x)^4 -60k^4 q^2 (-2-5h''(0)+12h''(0)^2 -2 h^{(4)}(0)))) \bigg) dx \ ,
 \end{multline}
 \begin{multline}
 \phi_7^2(v)=\frac{1}{64}\int_{-\infty}^v\bigg(144\phi_7^1(x)-76\phi_5^1(x)\fp(x)^2 +C_4(x)(72 P(v) -\fp(x)^3)+ k^4 q^2 \f(x)\fp(x)^2 h''(0) -\\
 24 k^4 q^2 \f(x) \fp(x)^2 h''(0)^2 + \phi_5^2(x)(-72(k^2 m+C_2(x))+28\fp(x)\fpp(x))+4k^4 q^2 \f(x)\fp(x)^2 h^{(4)}(0)\bigg) dx \ ,
\end{multline}
\begin{multline}
\phi_7^3(v)=\frac{1}{12}\int_{-\infty}^v\bigg(20 \phi_7^2(x) + 4 k^2\, q\, a_4(x) \fp(x)-10\phi_5^2(x)\fp(x)^2 + 4 k^2\, q\, a_4(x) \fp(x) h''(0)+\\
k^4\, q^2\, \f(x)^2 \fp(x)h''(0) - 4k^4\, q^2 \,\f(x)^2 \fp(x)h''(0)^2 +
k^4\, q^2\, \f(x)^2 \fp(x) h^{(4)}(0)\bigg) dx \ ,
 \end{multline}
 \begin{multline}
 \phi_7^4(v)=\frac{1}{24}\int_{-\infty}^v\bigg(24\phi_7^3(x)+6 C_6(x)\fp(x)+12 k^2\, q \,a_4(x) \fp(x) h''(0) + k^4\, q^2\, \f(x)^3h^{(4)}\bigg) dx \ .
\end{multline}

\subsubsection*{Order $\epsilon^8$:}

$\phi_8(z,v)=0$.
\begin{equation}
\bullet \,  g_8(z,v)=z\,C_8(v)+ \frac{1}{z} \big(z^8 g_8^0(v))+ z^7 g_8^1(v))+ z^6 g_8^2(v))+ z^5 g_8^3(v))+ z^4 g_8^4(v))+ z^3 g_8^5(v)\big) \ ,
\end{equation}

where
\begin{equation}
C_{8}(v)= \int_{-\infty}^{v} \frac{3}{2} \phi_7^{4}(x)\fp(x) dx \ ,
\end{equation}
\begin{multline}
g_8^0(v)=-\frac{1}{21}\left(\frac{9}{320} \k^2\, m\, P(v)^2 +\frac{9}{320}C_2(v)P(v)^2-\frac{1}{8} k^2 \, m\, \phi_5^0\fp(v) -\frac{1}{8} C_2(v) \phi_5^0(v)\fp(v) \right. \\
+ \frac{179}{1280} k^2\, m\, P(v) \fp(v)^3 + \frac{179}{1280} C_2(v)P(v)\fp(v)^3 +\frac{943}{61440} k^2\, m\, \fp(v)^6+\frac{943}{61440}C_2(v)\fp(v)^6 -\frac{21}{8}\phi_7^0(v) \fpp(v) \\
\left. -\frac{99}{649}P(v)^2 \fp(v)\fpp(v) -\frac{21}{64}\phi_5^0(v)\fp(v)^2\fpp(v)+\frac{237}{5120}P(v)\fp(v)^4\fpp(v)+\frac{147}{20480}\fp(v)^7\fpp(v) \right) \ ,
\end{multline}
\begin{multline}
g_8^1(v)=-\frac{1}{15} \left( \frac{5}{2}P(v)\phi_5^0(v)-\frac{1}{2}k^4\, q^2\, P(v)\fp(v)-\frac{3}{2}\phi_7^0(v)\fp(v)-\frac{43}{140}P(v)^2\fp(v)^2-\frac{13}{21}\phi_5^0(v)\fp(v)^3 \right. \\
\frac{107}{672}k^4\,q^2\,\fp(v)^4-\frac{53}{3360}P(v)\fp(v)^5+\frac{481}{40320}\fp(v)^8+\frac{4}{7}k^4\,q^2\,P(v)\fp(v)h''(0)+\frac{65}{336}k^4\, q^2\fp(v)^4h''(0)-\\
\left. \frac{9}{56}k^4\,q^2\fp(v)^4 h''(0)^2 -\frac{15}{7}\phi_7^6(v)\fpp(v)-\frac{3}{14}\phi_5^1(v)\fp(v)62\fpp(v)+\frac{3}{112}k^4\, q^2\fp(v)^4h^{(4)}(0) \right) \ ,
\end{multline}
\begin{multline}
g_8^2(v) = -\frac{1}{10} \left (2P(v)\phi_5^1(v)-\frac{1}{8}C_4(v)P(v)\fp(v)+\frac{1}{8}k^2\,m\, \phi_5^2(v)\fp(v)+\frac{1}{8}C_2(v)\phi_5^2(v)\fp(v) \right. \\
-\frac{5}{4}\phi_7^1(v)\fp(v)-\frac{23}{48}\phi_5^1(v)\fp(v)^3-\frac{13}{192}C_4(v)\fp(v)^4+\frac{1}{2}k^4\, q^2\, \f(v)P(v)h''(0)+\frac{77}{192}k^4\,q^2\f(v)\fp(v)^3h''(0)\\
\left. -\frac{5}{8}k^4\,q^2\, \f(v)\fp(v)^3 h''(0)^2-\frac{5}{3}\phi_7^5(v)\fpp(v)-\frac{5}{48}\phi_5^2(v)\fp(v)^2\fpp(v)+\frac{5}{48}k^4\,q^2\,\f(v)\fp(v)^3 h^{(4)}(0) \right) \ ,
\end{multline}

\begin{multline}
g_8^3(v) = -\frac{1}{6} \left(\frac{9}{5}P(v)\phi_5^2(v)-\phi_7^2(v)\fp(v)-\frac{2}{5}\phi_5^2(v)\fp(v)^3-\frac{9}{40}k^4\,q^2\,\f(v)^2\fp(v)^2h''(0) - \right. \\
 \frac{3}{5}k^4\,q^2\f(v)^2\fp(v)^2h''(0)^2+\frac{3}{10}k^2\,q a_4(v)\fp(v)^2(-3+2h''(0)) \\
\left. -\frac{6}{5}\phi_7^3(v)\fpp(v)+\frac{3}{20}k^4\,q^2\,\f(v)^2\fp(v)^2h^{(4)}(0) \right) \ ,
\end{multline}

\begin{multline}
g_8^4(v) = -\frac{1}{3} \left(-\frac{3}{4}\phi_7^3(v)\fp(v)-\frac{3}{16}C_6(v)\fp(v)^2+\frac{9}{8}k^2\,q\, a_4(v)\f(v)\fp(v)h''(0)-\frac{3}{4}\phi_7^4(v)\fpp(v) + \right. \\
\left. \frac{3}{332}k^4\,q^2\,\f(v)^2\fp(v)h^{(4)}(0) \right) \ ,
\end{multline}

\begin{multline}
g_8^5(v) = - \left(\frac{1}{2}a_4(v)^2-k^2\, q\, a_6(v)-\frac{1}{2}\phi_7^4(v)\fp(v)-\frac{1}{2}k^2\,q\,a_4(v)\f(v)^2h''(0)-\frac{1}{48}k^4\,q^2\,\f(v)^4h^{(4)}\right) \ .
\end{multline}

\begin{equation}
\bullet \,  f_8(z,v)= \frac{1}{z} \left(f_8^0(v)z^8+f_8^1(v)z^7+f_8^2(v)z^6+f_8^3(v)z^5+f_8^4(v)z^4\right) \ ,
\end{equation}

where

\begin{multline}
f_8^0(v)=-\frac{1}{112} \left(-15P(v)\phi_5^0(v)-7\phi_7^0(v)\fp(v)+\frac{39}{40}P(v)^2\fp(v)^2 \right. \\
\left. +\frac{2}{3}\phi_5^0(v)\fp(v)^3-\frac{11}{960}P(v)\fp(v)^5+\frac{\fp(v)^8}{92160} \right) \ ,
\end{multline}

\begin{equation}
f_8^1(v)=\frac{1}{84} \left(-12P(v)\phi_5^1(v)-\phi_7^1(v)\fp(v)+\frac{11}{20}\phi_5^4(v)\fp(v)^3 \right) \ ,
\end{equation}

\begin{equation}
f_8^2(v) = \frac{1}{60}\left(-9P(v)\phi_5^2(v)-5\phi_7^2(v)\fp(v)+\frac{7}{16}\phi_5^2(v)\fp(v)^3\right) \ ,
\end{equation}

\begin{equation}
f_8^3(v)=-\frac{1}{10}\phi_7^3(v)\fp(v) \ ,
\end{equation}

\begin{equation}
f_8^4(v)=-\frac{1}{8}\phi_7^4(v)\fp(v) \ .
\end{equation}

\begin{equation}
\bullet \, A_{v_8}(z,v) = \mu_8(v) +  A_{v_8}^0(v)z^7 + A_{v_8}^1(v)z^6 +A_{v_8}^2(v) z^5 + A_{v_8}^3(v) z^4 + A_{v_8}^4(v)z^3 + A_{v_8}^5(v) z^2 + A_{v_8}^6(v) z \ ,
\end{equation}

with

\begin{multline}
A_{v_8}^0(v) = -\frac{k^2\, q}{42} \left(-\frac{9}{10}P(v)^2-\phi_5^0(v)\fp(v)-\frac{19}{40}P(v)\fp(v)^3-\frac{17}{480}\fp(v)^6+3P(v)^2h''(0) \right.\\
+6 \phi_5^0(v)\fp(v)h''(0) +\frac{9}{4}P(v)\fp(v)^3h''(0)+\frac{1}{8}\fp(v)^6h''(0)-6 p(v)\fp(v)^3h''(0)^2-\frac{3}{8}\fp(v)^6h''(0)^2+\frac{3}{4}\fp(v)^6h''(0)^3 \\
\left. +P(v)\fp(v)^3h^{(4)}(0) +\frac{1}{16}\fp(v)^6h^{(4)}(0)-\frac{1}{4}\fp(v)^6h''(0)h^{(4)}(0)+\frac{1}{120}\fp(v)^6h^{(6)}(0) \right) \ ,
\end{multline}
\begin{multline}
A_{v_8}^1(v) = -\frac{k^2\, q}{30} \left(-\phi_5^1(v)\fp(v)+5 \f(v)\phi_5^0(v)h''(0)+5 \phi_5^1(v)\fp(v)h''(0)+\frac{5}{2}\f(v)P(v)\fp(v)^2h''(0)+ \right.\\
\frac{5}{24}\f(v)\fp(v)^5h''(0)-15\f(v)P(v)\fp(v)^2h''(0)^2-\frac{5}{4}\f(v)\fp(v)^5h''(0)^2+\frac{15}{4}\f(v)\fp(v)^5h''(0)^3+\\
\left. \frac{5}{2}\f(v)P(v)\fp(v)^2h^{(4)}(0)-\frac{5}{4}\f(v)\fp(v)^5h''(0)h^{(4)}(0)+\frac{1}{24}\f(v)\fp(v)^5h^{(6)}(0) \right) \ ,
\end{multline}

\begin{multline}
A_{v_8}^2(v) = -\frac{1}{20} \left(-a_4(v)P(v)\fp(v)-k^2\,\,\phi_5^2(v)\fp(v)-\frac{1}{6}a_4(v)\fp(v)^4+4k^2\,q\,\f(v)\phi_5^1(v)h''(0)+ \right. \\
4 a_4(v)P(v)\fp(v)h''(0)+4 k^2\,q\, \phi_5^2(v)\fp(v)h''(0)+4k^2\,q\,\phi_5^2(v)\fp(v)h''(0)+\frac{1}{2}a_4(v)\fp(v)^4h''(0)-\\
10 k^2\,q\,\f(v)^2 P(v)\fp(v)h''(0)^2-a_4(v)\fp(v)^4h''(0)^2-\frac{5}{4}k^2\, q\,\f(v)^2\fp(v)^4h''(0)^2+7k^2\,q\,\f(v)^2\fp(v)^4h''(0)^3+\\
2k^2\,\,\f(v)^2P(v)\fp(v)h^{(4)}(0)+\frac{1}{6}a_4(v)\fp(v)^4h^{(4)}(0)-\frac{29}{12}k^2\,q\,\f(v)^2\fp(v)^4h''(0)h^{(4)}(0)\\
\left. +\frac{1}{12}k^2\,q\,\fp(v)^2\fp(v)^4h^{(6)}(0) \right) \ ,
\end{multline}
\begin{multline}
A_{v_8}^3(v) = -\frac{1}{12} \left(3a_4(v)\f(v)P(v)h''(0)+ 3k^2\,q\, \f(v)\phi_5^2(v)h''(0)+\frac{3}{4}a_4(v)\f(v)\fp(v)^3h''(0)- \right. \\
\frac{3}{2}k^2\, q\, \f(v)^3P(v)h''(0)^2-3a_4(v)\f(v)\fp(v)^3h''(0)^2-\frac{3}{8}k^2\,\,q\f(v)^3\fp(v)^3h''(0)^2+6\,k^2\,q\, \f(v)^3 \fp(v)^3h''(0)^3+\\
\frac{1}{2}k^2\, q\, \f(v)^3 P(v)h^{(4)}(0)+\frac{1}{2}a_4(v)\f(v)\fp(v)^3h^{(4)}(0)+\frac{1}{8}k^2\,q\, \f(v)^3\fp(v)^3h^{(4)}(0)-\\
\left. \frac{9}{4}k^2\, q\, \f(v)^3 \fp(v)^3h''(0)h^{(4)}(0)+\frac{1}{12}k^2\, q\, \f(v)^3\fp(v)^3 h^{(6)}(0) \right) \ ,
\end{multline}
\begin{multline}
A_{v_8}^4(v) = -\frac{1}{6} \left(-\frac{1}{2} A_{v_6}^4(v)\fp(v)^2+A_{v_6}^4(v)\fp(v)^2h''(0)-\frac{5}{2}a_4(v)\fp(v)^2\fp(v)^2h''(0)^2 \right. \\
 +\frac{9}{4}k^2\,q\,\fp(v)^4\fp(v)^2h''(0)^3 +\frac{1}{2}a_4(v)\fp(v)^2\fp(v)^2h^{(4)}(0) -\frac{23}{24}k^2\,q\, \f(v)^2h''(0)h^{(4)}(0) \\
\left. +\frac{1}{24}k^2\,q\, \f(v)^4\fp(v)^2h^{(6)}(0) \right) \ ,
\end{multline}
\begin{multline}
A_{v_8}^5(v) = -\frac{1}{2} \left(A_{v_6}^4(v)\f(v)\fp(v)h''(0)-\frac{1}{2}a_4(v)\f(v)^3\fp(v)h''(0)^2+\frac{1}{4}k^2\,\,q \, \f(v)^5\fp(v)h''(0)^3+ \right.\\
\left. \frac{1}{6}a_4(v)\f(v)^3\fp(v)h^{(4)}(0)-\frac{1}{8}k^2\,q\, \f(v)^5\fp(v)h''(0)h^{(4)}(0)+\frac{1}{120}k^2\, q\, \f(v)^5\fp(v)h^{(6)}(0) \right) \ ,
\end{multline}

\begin{multline}
A_{v_8}^6(v)=-\frac{1}{120}\int_{-\infty}^v\f(x)\fp(x) \left(120 A_{v_6}^4(x)h''(0)-60 a_4(x)\f(x)^2h''(x)^2+30k^2\,q\, \f(x)^4h''(0)^3 + \right. \\
\left. 20 a_4(x)\f(x)^2h^{(4)}(0)-15k^2\,q\, \f(x)^4h''(0)h^{(4)}(0)+k^2\,q\,\f(x)^4h^{(6)}(0) \right) dx \ .
\end{multline}



\end{document}